\documentclass[preprint,showpacs,preprintnumbers,superscriptaddress,amsmath,
amssymb]{revtex4}

\usepackage{graphicx}
\usepackage{dcolumn}
\usepackage{bm,morefloats,color}

\usepackage{amssymb,bm}

\def\Eqref#1{Eq.~(\ref{#1})}

\def\Eq#1{\begin{equation} #1 \end{equation}}
\def\Eqr#1{\begin{eqnarray} #1 \end{eqnarray}}
\def\Eqrsubl#1#2{\begin{subequations}\label{#1}\Eqr{#2}\end{subequations}}

\newcommand{\nn}{\nonumber}
\newcommand{\pd}{\partial}
\newcommand{\vect}[1]{\!\!\!\mbox{ \boldmath $#1$}}
\newcommand{\bea}{\begin{eqnarray}}
\newcommand{\eea}{\end{eqnarray}}

\def\Xsp{{\rm X}}

\def\Ysp{{\rm Y}}
\def\Zsp{{\rm Z}}
\def\X5sp{{\rm X}_5}
\def\Y3sp{{\rm Y}_3}
\def\Z3sp{{\rm Z}_3}

\def\Msp{{\rm M}}
\def\Nsp{{\rm N}}
\def\lap{{\triangle}}
\def\e{{\rm e}}

\begin{document}


\title{Dynamical brane with angles : Collision of the universes}

\baselineskip 14pt
\author{Kei-ichi Maeda}%
\affiliation{%
Department of Physics, Waseda University, Okubo 3-4-1,
Shinjuku, Tokyo 169-8555, Japan}%

\author{Kunihito Uzawa}
\affiliation{%
Department of Physics, Kinki University,
Higashi-Osaka, Osaka 577-8502, Japan
}%

\date{\today}

\begin{abstract}
\baselineskip 14pt
We present the time-dependent solutions corresponding to 
the dynamical D-brane with angles in ten-dimensional type II 
supergravity theories. 
Our solutions with angles are different from 
the known dynamical intersecting brane solutions 
in supergravity theories. 
Because of our ansatz for fields,
all warp factors in the solutions can depend on time.  
Applying these solutions,
we construct cosmological models from those solutions 
by smearing some dimensions and compactifying the internal space.
We find the Friedmann-Lema\^itre-Robertson-Walker (FLRW)
 cosmological solutions with power-law expansion.
We also discuss the dynamics of branes based on these solutions. 
When the spacetime is contracting in ten dimensions, 
each brane approaches the others as the time evolves.
However, for D$p$-brane ($p\le 7$) without smearing branes,  
a singularity appears before branes collide.
 In contrast, the D6-D8 brane system or the smeared D$(p-2)$-D$p$ brane 
system with one uncompactified extra dimension
can provide an example of colliding branes 
(and  collision of the universes), if they have the same charges.
\end{abstract}

\maketitle

\baselineskip 18pt

\section{Introduction}
\label{sec:introduction}
Lately, many dynamical solutions of $p$-brane system in 
different dimensions have been studied using various 
approaches based on string theory \cite{Gibbons:2005rt, Chen:2005jp, 
Kodama:2005fz, 
Binetruy:2007tu, Maeda:2009tq, Maeda:2009zi, Gibbons:2009dr, Maeda:2009ds, 
Maeda:2010yk, Maeda:2010ja, Minamitsuji:2010fp, Maeda:2010aj, 
Minamitsuji:2010kb, Minamitsuji:2010uz, Maeda:2011sh, Minamitsuji:2011jt}. 
These solutions have varying
applications and shed light on many different aspects of dynamics of the 
higher-dimensional spacetime. 

In a brane world scenario
\cite{brane_world0,brane_world1,brane_world2,brane_world3}, 
the brane dynamics is very important.
If we construct a cosmological scenario based on a 
brane world scenario from a fundamental unified theory,
we may find a brane inflation model\cite{brane_inflation,brane_collision}
or an alternative model such as a cyclic universe\cite{cyclic_universe}
in the early stage of the universe.
However, in most of those models, a probe test brane is assumed 
and the dynamics of our universe is discussed in a lower-dimensional
 effective theory. 
No back reaction is taken into account.
Since the existence of branes cause inhomegeneity of spacetime,
 a simple truncation of extra dimensions for  an effective theory 
 may lead us to a wrong answer\cite{Kodama:2005fz,Binetruy:2007tu, 
 Kodama:2005cz}, except for 
 Kaluza-Klein comcatification. 
We may have to discuss such a dynamics in the original higher dimensions.

The purpose of the present paper 
is to discuss a brane dynamics by use of exact solutions in higher dimensions.
In order to find appropriate solutions, we adopt the most classical riddles 
of higher-dimensional dynamics such as intersection, T-duality, 
and the relation, if any, between them in a situation as simple 
as possible while still more or less tractable.
For this aim, we explore a higher-dimensional time-dependent model that 
is a relatively close analog of ordinary supergravity theories. 
Such a model is the $p$-brane model with a $B$-field 
\cite{Breckenridge:1996tt, Youm:1997hw, DiVecchia:1999uf, Di Vecchia:1999rh, 
Di Vecchia:1999fx, Maldacena:1999mh, Lu:1999rm, Lu:2000ys, 
Cai:1999aw, Cai:2000hn, Cai:2000yk}. 
This model is possible to exhibit time dependence where all harmonic 
functions in the metric depend on time. 
In the intersecting brane system, the warp factors arise from field 
strengths. Then the dynamics of a system composed of $n$ branes can be 
characterized by $n$ warp factors arising from $n$ field strengths. 
Unfortunately, since we have ever found that among these warp factors for 
M-branes and D-branes, only one function can depend on time 
\cite{Binetruy:2007tu, Maeda:2009tq, Minamitsuji:2010kb, Minamitsuji:2010uz}, 
there are little-known solutions in which all harmonic 
functions depend on time for M-branes and D-branes.
These are some of the main properties that we would like to understand 
in cosmological solutions.
Many other interesting models contain cosmological solutions, 
as a result of which they are not such close relatives of 
supergavity theories if all warp factors in the metric depend on time 
\cite{Minamitsuji:2010uz}.
Another important property that the $p$-brane model 
is believed to share in common with ordinary Kaluza-Klein compactification 
 is a limit of cosmic time  
in which the time dependence in the warp factor is the 
dominant contribution, and the effects of field strengths 
vanish. For these cosmological $p$-brane solutions, 
though there are realistic cosmological model in the four-dimensional 
effective theory \cite{Gibbons:2009dr}, nobody knows the $p$-brane solution 
which exhibits an accelerating expansion of our universe 
in the viewpoint of original higher-dimensional supergravity 
\cite{Binetruy:2007tu, Maeda:2009tq, Minamitsuji:2010kb, Minamitsuji:2010uz,
Minamitsuji:2011jt}.
Understanding of this result, perhaps via a new kind of 
ansatz for fields, is probably well out of reach with present methods, 
but may offer the best long term hope of a much better understanding 
of cosmological evolutions than we now possess. Note that ten-dimensional  
string theory is believed to have the cosmological solution 
with accelerating expansion of our universe in the four-dimensional 
effective theory. This fact is important in the 
cosmological model of string theory or M-theory \cite{Kachru:2003aw, 
Kachru:2003sx, Silverstein:2007ac, Gibbons:2009dr}. 

To study the dynamics of the $p$-brane, 
we follow a path that has been followed for 
a variety of D-brane in ten-dimensional type II 
supergravity models: 
we construct a configuration of dynamical branes with $B$-field 
in a supergravity theory that realizes the string theory of 
interest at low energies, and then we find the cosmological 
solutions using the T duality map between the type IIA and 
type IIB superstring theories. 
A method of doing this in the case at hand will be described 
 in the sections \ref{sec:D3D5} and \ref{sec:Dp}.  
We establish by a geometrical argument a new result which has not 
been guessed previously: the two kinds of metric functions 
 depend on time as well as the coordinates of the 
transverse space to the intersecting brane 
for the D$(p-2)$- and D$p$-branes in 
the supergravity theory.
A dependence of angles in the ten-dimensional metric 
is obtained via  the T duality map. 
These dynamical solutions are  a
straightforward generalization of the bound state of a static 
D$(p-2)$- and D$p$-branes system with a dilaton coupling 
\cite{Breckenridge:1996tt, Cai:1999aw, Cai:2000hn, Cai:2000yk}. 
We consider in detail the 
construction yielding the D3-D5 brane. We also provide 
the brief discussions for other D$(p-2)$-D$p$ brane system
 in section \ref{sec:Dp}.
In section \ref{sec:cosmology}, we describe how our universe  
could be represented in the present formulation 
via an appropriate compactification
\cite{Binetruy:2007tu, Maeda:2009zi, Minamitsuji:2010kb, Minamitsuji:2010uz}.
We  show that there exist no accelerating expansion of 
our universe, although the conventional power-law expansion of the universe 
is possible. 
To illustrate this, we construct 
cosmological models for the D$(p-2)$-D$p$ brane system, which is
 relevant to the ordinary ten-dimensional type II string theory. 
We give the classification of the D$(p-2)$-D$p$ 
 brane system and the application to cosmology. 
We then discuss the dynamics of two D$p$ branes with smeared D$3$ branes
(or the dynamics of two universes) 
 in sec \ref{sec:collision}. 
If there exists one uncompactified extra dimension (D6-D8 brane system
or smeared D$(p-2)$-D$p$ brane systems $(p\leq 7)$) and two brane systems
have the same charges, the solution
describes a collision of two brane systems (or two universes),,  
which is similar to the result in \cite{Gibbons:2005rt, Maeda:2010aj}. 
 Section \ref{sec:discussions} is devoted to conclusion and remarks.

\section{Dynamical solution of the D3-D5 brane with angle}
\label{sec:D3D5}

We discuss the dynamical solutions for 
D3-D5 brane system in the string theory. 
The starting point is a D5-brane which carries an electric charge of 
the 7-form field strength. 
Then its Hodge dual gives the magnetic 3-form 
field strength, which  2-form potential is easily expressed by 
the coordinates of the transverse space of the D5-brane. 

For the D3-D5 brane system,  
the equations of motion of the ten-dimensional type IIB theory 
in the Einstein frame are written as 
\Eqrsubl{IIB:equations:Eq}{
&&R_{MN}=\frac{1}{2}\pd_M\phi \pd_N \phi
+\frac{1}{2\cdot 3!}\e^{\phi} 
\left(3F_{MAB} {F_N}^{AB}-\frac{1}{4}g_{MN} F_{(3)}^2\right)\nn\\
&&~~~~~~+\frac{1}{2\cdot 3!}\e^{-\phi} 
\left(3H_{MAB} {H_N}^{AB}
-\frac{1}{4}g_{MN} H_{(3)}^2\right)
+\frac{1}{4\cdot 4!}F_{MABCD} {F_N}^{ABCD}\,,
   \label{IIB:Einstein:Eq}\\
&&d\ast d\phi=\frac{1}{2\cdot 3!}
\e^{\phi}F_{(3)}\wedge \ast F_{(3)}
-\frac{1}{2\cdot 3!}\e^{-\phi}H_{(3)}\wedge \ast H_{(3)}\,,
   \label{IIB:scalar:Eq}\\
&&d\left(\e^{\phi}\ast F_{(3)}\right)+H_{(3)}\wedge\ast F_{(5)}=0\,,
   \label{IIB:F:Eq}\\
&&d\left(\e^{-\phi}\ast H_{(3)}\right)
 +\ast F_{(5)}\wedge F_{(3)}=0\,,\\
&&F_{(5)}=\pm\ast F_{(5)}\,,
   \label{IIB:H:Eq}
}
where $\ast$ is the Hodge operator in the ten-dimensional spacetime, and 
we define 
\Eqrsubl{IIB:strength:Eq}{
F_{(3)}&=&dC_{(2)}\,,\\
H_{(3)}&=&dB_{(2)}\,,\\
F_{(5)}&=&dC_{(4)}+\frac{1}{2}\left(C_{(2)}\wedge H_{(3)}-
B_{(2)}\wedge F_{(3)}\right)\,.
}
Here $B_{(2)}$, $C_{(2)}$ and $C_{(4)}$ are the NS 2-form, 
RR 2-form, and RR 4-form, respectively. 

To solve the field equations, we assume 
a brane configuration shown in the following Table \ref{table_D3D5}:
\begin{table}[h]
{\scriptsize
\begin{center}
\begin{tabular}{|c||c|c|c|c|c|c|c|c|c|c|}
\hline
$A$&0&1&2&3&4&5&6&7&8&9\\
\hline
$X^A$&\multicolumn{4}{c}{$x^\mu$}&\multicolumn{2}{|c}{$y^i$}&
\multicolumn{4}{|c|}{$z^a$}
\\
\hline
\hline
D3 & {\normalsize $\circ$} & {\normalsize $\circ$}
 & {\normalsize $\circ$} & {\normalsize $\circ$} &&&&&& \\
\hline
D5 & {\normalsize $\circ$} & {\normalsize $\circ$} &
{\normalsize $\circ$} & {\normalsize $\circ$} &
{\normalsize $\circ$} & {\normalsize $\circ$}
&& & & \\ 
\hline
\end{tabular}
\end{center}
}
\caption{\baselineskip 14pt
Brane configuration for a D3-D5 brane system.}
\label{table_D3D5}
\end{table}

Then we assume the ten-dimensional metric as
\Eqr{
ds^2&=& h^{1/2}(x, z)h_\theta^{1/4}(x, z)\left[h^{-1}(x, z)q_{\mu\nu}(\Xsp)
    dx^{\mu}dx^{\nu}\right.\nn\\
     & &\left.+h_\theta^{-1}(x, z)\gamma_{ij}(\Ysp)dy^idy^j
     +u_{ab}(\Zsp)dz^adz^b\right],
   \label{D3D5:metric:Eq}
     }
where $q_{\mu\nu}$, $\gamma_{ij}$, and $u_{ab}$ are
the  metric of the four-dimensional spacetime $\Xsp$, 
that of the two-dimensional space $\Ysp$,
and that of the four-dimensional space  $\Zsp$, which depend  
 only on the four-dimensional coordinates $x^{\mu}$, 
 on the two-dimensional ones $y^i$, 
and  on the four-dimensional ones $z^a$, respectively.
The function $h_\theta$, which depends on $x^{\mu}$ and $z^a$,
 is given by
\Eq{
h_\theta(x, z)=1+ \cos^2\theta\left[h(x, z)-1\right]
\,,
   \label{D3D5:f:Eq}
 }
where $\theta$ is an angle parameter and the warp factor $h(x, z)$ is 
a function to be solved.
The metric form (\ref{D3D5:metric:Eq}) is
a straightforward generalization of the case of a static bound state 
of D3-D5-brane system with a dilaton coupling \cite{Maldacena:1999mh}.
This ansatz denotes that 
the D3-brane is set in array parallelly on the D5-brane in order to 
smear the space $\Ysp$.
 We  find this configuration 
via the T duality map between the type IIA and 
type IIB superstring theories, which we describe later. 

Furthermore, we assume that the dilaton field $\phi$ and 
the gauge potentials are given by
\Eqrsubl{D3D5:fields:Eq}{
\e^{\phi}&=&h_\theta^{-1/2},\\
C_{(2)}&=&\cos\theta\,\omega_{(2)}\,,\\
B_{(2)}&=&\tan\theta\,\left(h_\theta^{-1}-1\right)\Omega(\Ysp)\,,\\
C_{(4)}&=&\omega_{(4)}\pm\sin\theta h^{-1}\Omega(\Xsp)\,,
}
where $\Omega(\Xsp)$ and $\Omega(\Ysp)$ denote the volume forms,
defined by 
\Eqrsubl{D3D5:volume:Eq}{
\Omega(\Xsp)&=&\sqrt{-q}\,dt\wedge dx^1\wedge dx^2\wedge dx^3,
    \label{D3D5:volume-x:Eq}\\
\Omega(\Ysp)&=&\sqrt{\gamma}\,dy^1\wedge dy^2,
   \label{D3D5:volume-y:Eq}
}
and the 2-form $\omega_{(2)}$ and the 4-form $\omega_{(4)}$ satisfy 
\Eqrsubl{D3D5:omega:Eq}{
d\omega_{(2)}&=&\mp \,\pd_a h\ast_{\Zsp}\left(dz^a\right),\\
\omega_{(4)}&=&-\frac{1}{2}\sin\theta
\left(h_\theta^{-1}+1\right)\Omega(\Ysp)\wedge \omega_{(2)}\,,
}
respectively. Here $\ast_{\Zsp}$ denotes the Hodge operator on $\Zsp$. 

Let us first consider the Einstein equations 
(\ref{IIB:Einstein:Eq}). 
Using the assumptions (\ref{D3D5:metric:Eq}) and 
(\ref{D3D5:fields:Eq}), the Einstein equations are reduced to
\Eqrsubl{D3D5:cEinstein:Eq}{
&&R_{\mu\nu}(\Xsp)-h^{-1}D_{\mu}D_{\nu}h
+\frac{1}{8}q_{\mu\nu}h^{-1}\left(2-\cos^2\theta hh_\theta^{-1}\right)
\left(\Box_{\Xsp}h+h^{-1}\lap_{\Zsp}h\right)=0\,,
 \label{D3D5:cEinstein-mn:Eq}\\
&&h^{-1}\pd_{\mu}\pd_a h=0\,,
 \label{D3D5:cEinstein-mz:Eq}\\ 
&&R_{ij}(\Ysp)-\frac{1}{4}\gamma_{ij}h_\theta^{-1}
\left(2-3\cos^2\theta hh_\theta^{-1}\right)
\left(\Box_{\Xsp}h+h^{-1}\lap_{\Zsp}h\right)=0\,,
 \label{D3D5:cEinstein-ij:Eq}\\
&&R_{ab}(\Zsp)-\frac{1}{8}u_{ab}\left(2+\cos^2\theta hh_\theta^{-1}\right)
\left(\Box_{\Xsp}h+h^{-1}\lap_{\Zsp}h\right)=0\,,
 \label{D3D5:cEinstein-ab:Eq}
}
where $D_{\mu}$ is the covariant derivative with respective to 
the metric $q_{\mu\nu}(\Xsp)$, and $\Box_{\Xsp}$, $\triangle_{\Zsp}$ are 
the Laplace operator on the space X, $\Zsp$, and $R_{\mu\nu}(\Xsp)$, 
$R_{ij}(\Ysp)$, and $R_{ab}(\Zsp)$ are the Ricci tensors of the metric 
$q_{\mu\nu}(\Xsp)$, $\gamma_{ij}(\Ysp)$, and $u_{ab}(\Zsp)$, respectively.

From \Eqref{D3D5:cEinstein-mz:Eq}, 
the warp factor $h$ must be in the form
\Eq{
h(x, z)= h_0(x)+h_1(z)\,.
  \label{D3D5:warp:Eq}
}
With this form of $h$, the other components of
the Einstein equations (\ref{D3D5:cEinstein:Eq}) 
are rewritten as
\Eqrsubl{D3D5:c1Einstein:Eq}{
&&R_{\mu\nu}(\Xsp)-h^{-1}D_{\mu}D_{\nu}h_0
+\frac{1}{8}q_{\mu\nu}h^{-1}\left(2-\cos^2\theta hh_\theta^{-1}\right)
\left(\Box_{\Xsp}h_0+h^{-1}\lap_{\Zsp}h_1\right)=0\,,
 \label{D3D5:c1Einstein-mn:Eq}\\
&&R_{ij}(\Ysp)-\frac{1}{4}\gamma_{ij}h_\theta^{-1}
\left(2-3\cos^2\theta hh_\theta^{-1}\right)
\left(\Box_{\Xsp}h_0+h^{-1}\lap_{\Zsp}h_1\right)=0\,,
 \label{D3D5:c1Einstein-ij:Eq}\\
&&R_{ab}(\Zsp)-\frac{1}{8}u_{ab}\left(2+\cos^2\theta hh_\theta^{-1}\right)
\left(\Box_{\Xsp}h_0+h^{-1}\lap_{\Zsp}h_1\right)=0\,.
 \label{D3D5:c1Einstein-zz:Eq}
}

Let us next consider gauge fields. 
In terms of the ansatz \eqref{D3D5:fields:Eq} 
for fields, 
the field equations for $F_{(3)}$ and $F_{(5)}$ are automatically 
satisfied. 
As a result, the equation of motion for the gauge field $H_{(3)}$ gives 
\Eqr{
&&\sin 2\theta\left(\Box_{\Xsp}h_0+h^{-1}\lap_{\Zsp}h_1\right)
\,\Omega(\Xsp) \wedge \Omega(\Zsp)=0\,,
   \label{D3D5:H3:Eq}
 }
where we have used \eqref{D3D5:warp:Eq}. 
$\Omega(\Zsp)$ denotes the volume 4-form,
\Eq{
\Omega(\Zsp)=\sqrt{u}\,dz^1\wedge \cdots \wedge dz^4.
    \label{D3D5:volume-z:Eq}\
}
The equation of motion for gauge field $H_{(3)}$ is thus reduced to 
\Eq{
\Box_{\Xsp}h_0=0,~~~~~\lap_{\Zsp} h_1=0\,.
   \label{D3D5:H3-2:Eq}
}

Next we consider the dilaton field equation 
\eqref{IIB:scalar:Eq}.
Substituting Eqs. \eqref{D3D5:metric:Eq} and 
\eqref{D3D5:fields:Eq}  
into the equation of motion  
\eqref{IIB:scalar:Eq}, we find 
\Eqr{
h^{-3/2}h_\theta^{-5/4}
\left(\Box_{\Xsp}h_0+h^{-1}\lap_{\Zsp}h_1\right)=0\,,
   \label{D3D5:scalar2:Eq}
}
where we used Eqs.~(\ref{D3D5:f:Eq}) and (\ref{D3D5:warp:Eq}). 
Thus, the warp factor $h$ should
satisfy the equations
\Eq{
\Box_{\Xsp}h_0=0\,, ~~~ \lap_{\Zsp} h_1=0\,.
   \label{D3D5:warp3:Eq}
}
Hence if one assumes 
\Eqrsubl{D3D5:Einstein equations:Eq}{
&&R_{\mu\nu}(\Xsp)=0\,,
   \label{D3D5:Ricci tensor mu:Eq}\\
&&R_{ij}(\Ysp)=0\,,
   \label{D3D5:Ricci tensor ij:Eq}\\
&&R_{ab}(\Zsp)=0\,,
   \label{D3D5:Ricci tensor pq:Eq}\\ 
&&D_{\mu}D_{\nu}h_0=0\,,   \label{D3D5:warp factor h0:Eq}\\
&&\lap_{\Zsp}h_1=0\,,
   \label{D3D5:warp factor h1:Eq}
 }      
all equations are solved with 
the additional conditions 
\Eqrsubl{D3D5:additional conditions}{
&&h(x, z)=h_0(x)+h_1(z)\,,
   \label{D3D5:warp factor h:Eq}\\
&&h_\theta(x, z)=1+\cos^2\theta(h-1)\,.
   \label{D3D5:warp factor h5:Eq} 
 }      

To see the solutions more explicitly, 
let us consider the case of  $q_{\mu\nu}=\eta_{\mu\nu}$ 
and $u_{ab}=\delta_{ab}$, 
where $\eta_{\mu\nu}$ is the four-dimensional Minkowski metric, and 
$\delta_{ab}$ are the four-dimensional flat Euclidean metric.
In this case, a general solution for the warp factor $h$ 
is obtained as
\Eqr{ 
h(x, z)=c_{\mu} x^{\mu}+\tilde{c}+\sum_l\frac{M_l}{|z^a-z^a_l|^2}\,,
 \label{D3D5:h:Eq}
}
where $c_{\mu}$, $\tilde{c}$, $M_l$ and $z_l$ are 
integration constants.
If $c_0\neq 0$, the solution (\ref{D3D5:h:Eq}) 
 depends on time $t$.

Near any brane,
which we assume to be located at the origin 
without loss of generality,
writing 
\Eq{
\gamma_{ij}=\delta_{ij}\,,~~~\delta_{ab}dz^adz^b=dr^2+r^2d\Omega_3^2\,, 
}
where $d\Omega_3^2$ is the line element of three-dimensional sphere,
we find that the warp factor $h$ is approximated by 
\Eq{
h(x, r)\approx \left(\frac{r_0}{r}\right)^2\,,
}
where $r_0$ is a constant, as $r\rightarrow 0$. 
Then, the metric \eqref{D3D5:metric:Eq} near the brane reads
\Eqr{
\hspace{-0.7cm}
ds^2=\cos^{1/2}\theta\left(\frac{r_0}{r}\right)^{-1/2}
\left(\eta_{\mu\nu}dx^{\mu}dx^{\nu}+\cos^{-2}\theta\delta_{ij}dy^idy^j
+\frac{r_0^2}{r^2}dr^2+r_0^2d\Omega_3^2\right)\,,
   \label{D3D5:metric2:Eq}
}
which is static, 
while the dilaton field near the brane is given by
\Eq{
\e^{\phi}\approx \left(\cos\theta\right)^{-1}
\left(\frac{r}{r_0}\right)\,.
    \label{D3D5:dilaton2:Eq}
}
Ten-dimensional metric and dilaton field are static near any brane, and 
the spacetime is described by 
a warped  geometry of ${\rm AdS}_7\times {\rm S}^3$.

Now we show how to obtain 
the solution (\ref{D3D5:warp:Eq}) via the T-duality.
We start from the 
dynamical D4-brane solution in the string frame in the type IIA theory;
\Eqrsubl{D4:solution:Eq}{
ds^2_{(\rm A)}&=& h^{1/2}\left[h^{-1}\left(-dt^2+d\tilde{x}^2
     +\delta_{mn}dv^mdv^n\right)+d\tilde{y}^2
     +\delta_{ab}(\Zsp)dz^adz^b\right]\,,
   \label{D4:metric:Eq}\\
C_{(3)}&=&\omega_{(3)}\,,\\
\e^{2\phi_{(\rm A)}}&=&h^{1/2}\,,
}
where $(\tilde{x}, v^m)$  are the world volume coordinates of the D4-brane, 
and $(\tilde{y}, z^a)$ are the coordinates of the 
transverse space. 
$\delta_{mn}$ and $\delta_{ab}$ are the three-, and five-dimensional 
Euclidean metrics.
The 
warp factor $h$ is given by
\Eq{
h=c_0t+c_mv^m+\tilde c+\sum_{l}\frac{M_l}{r_l^3}\,,
 \label{D4:h:Eq}
}
where $c_0$, $c_m$, $\tilde c$ and $M_l$ 
are arbitrary constants, and $r_l$ is defined by 
\Eq{
r_l^2=(\tilde{y}-\tilde{y}_l)^2+\sum_{a=1}^4|z^a-z_l^a|^2\,.
}
Here $(\tilde{y}_A,z_l^a)$ is the position of the $l$-th brane.
The 3-form $\omega_{(3)}$ should satisfy the relation 
\Eq{
d\omega_{(3)}=\pm\pd_ah\,d\tilde{y}\wedge\ast_{\Zsp}dz^a\,.
}
Now we delocalize the D4-brane in one of the transverse 
directions where we have singled out one of the transverse coordinate 
$\tilde{y}$. 
Since D4 brane is smeared out in the $\tilde{y}$ direction, 
the number of transverse dimensions to D4-brane becomes effectively four. 
Then, the function $h$ in Eq. \eqref{D4:h:Eq} is replaced as 
\Eq{
h(t, z)=c_0t+c_mv^m+\tilde c+\sum_{l}\frac{M_l}{|z^a-z^a_l|^2}\,.
 \label{D4:h2:Eq}
}
Let us consider some rotation in the $(\tilde{x}, \tilde{y})$ 
plane of the ten-dimensional metric \eqref{D4:metric:Eq}
by an angle $\theta$ such that
\Eqr{
\left(
\begin{array}{c}
 x \\
y
\end{array} \right)
=\left(
\begin{array}{cc}
 \cos\theta &-\sin\theta \\
 \sin\theta &\cos\theta 
\end{array} \right)
\left(
\begin{array}{c}
 \tilde{x}\\
 \tilde{y}
\end{array} \right).
 \label{p:transform:Eq}
}
Under the rotation \eqref{p:transform:Eq}, the ten-dimensional metric 
\eqref{D4:metric:Eq} becomes 
\Eqrsubl{D4:solution2:Eq}{
ds^2_{(\rm A)}&=& h^{1/2}
\left[h^{-1}\left(-dt^2+\delta_{mn}dv^mdv^n\right)
     +\left(\sin^2\theta+h^{-1}\cos^2\theta\right)dx^2\right.\nn\\
    &&\left.+\left(\cos^2\theta+h^{-1}\sin^2\theta\right)dy^2
    -2\cos\theta\sin\theta(1-h^{-1})
    +\delta_{ab}(\Zsp)dz^adz^b\right],
   \label{D4:metric2:Eq}\\
C_{(3)}&=&\omega_{(3)}\,,\\
\e^{2\phi_{(\rm A)}}&=&h^{1/2}\,,
} 
where the 3-form $\omega_{(3)}$ have to satisfy the relation 
\Eq{
d\omega_{(3)}=\pm\pd_ah\,\left(-\sin\theta dx+\cos\theta dy\right)
\wedge\ast_{\Zsp}dz^a\,.
}
Here $\ast_{\Zsp}$ denotes the Hodge operator on $\Zsp$. 
Now we will obtain the dynamical solution of  
a D3-D5 brane after we apply T duality in the $y$ direction of the 
ten-dimensional spacetime (\ref{D4:metric:Eq}).
The ten-dimensional T duality map from the type IIA theory to type IIB 
theory is given by \cite{Bergshoeff:1995as, Breckenridge:1997ar}
\Eqr{
&&g^{(\rm B)}_{yy}=\frac{1}{g^{(\rm A)}_{yy}}\,,~~~~
g^{(\rm B)}_{\mu\nu}=g^{(\rm A)}_{\mu\nu}
-\frac{g^{(\rm A)}_{y\mu}g^{(\rm A)}_{y\nu}
-B^{(\rm A)}_{y\mu}B^{(\rm A)}_{y\nu}}
{g^{(\rm A)}_{yy}}\,,~~~~
g^{(\rm B)}_{y\mu}=-\frac{B^{(\rm A)}_{y\mu}}{g^{(\rm A)}_{yy}}\,,\nn\\
&&\e^{2\phi_{(\rm B)}}=\frac{\e^{2\phi_{(\rm A)}}}{g^{(\rm A)}_{yy}}\,,~~~~
B^{(\rm B)}_{\mu\nu}=B^{(\rm A)}_{\mu\nu}
+2\frac{g^{(\rm A)}_{y[\mu}\,B^{(\rm A)}_{\nu]y}}{g^{(\rm A)}_{yy}},~~~~
B^{(\rm B)}_{y\mu}=-\frac{g^{(\rm A)}_{y\mu}}{g^{(\rm A)}_{yy}},\nn\\
&&C_{\mu\nu}=C_{\mu\nu y}-2C_{[\mu}B_{\nu]y}^{(\rm A)}
+2\frac{g^{(\rm A)}_{y[\mu}\,B^{(\rm A)}_{\nu]y}C_y}{g^{(\rm A)}_{yy}}\,,
~~~~C_{y\mu}=C_{\mu}
-\frac{A^{(\rm A)}_{y}\,g^{(\rm A)}_{y\mu}}{g^{(\rm A)}_{yy}}\,,\nn\\
&&C_{\mu\nu\rho y}=C_{\mu\nu\rho}-\frac{3}{2}\left(
C_{[\mu}\,B^{(\rm A)}_{\nu\rho]}
-\frac{g^{(\rm A)}_{y[\mu}B^{(\rm A)}_{\nu\rho]}\,C_y}
{g^{(\rm A)}_{yy}}
+\frac{g^{(\rm A)}_{y[\mu}\,C_{\nu\rho]y}}
{g^{(\rm A)}_{yy}}\right)\,,~~~~C_{(0)}=-C_y\,,
   \label{D4:duality:Eq}
}
where $y$ is the coordinate to which the T dualization is applied, and 
$\mu$, $\nu$, $\rho$ denote the coordinates other than $y$.
In terms of the T-duality map \eqref{D4:duality:Eq}, the solution 
\eqref{D4:solution2:Eq} becomes  
\Eqrsubl{D3D5:t-solution:Eq}{
ds^2_{(\rm B)}&=& h^{1/2}\left[h^{-1}\left(-dt^2+\delta_{mn}dv^mdv^n\right)
   +h_\theta^{-1}\left(dx^2+dy^2\right)
     +\delta_{ab}dz^adz^b\right]\,,
   \label{D3D5:t-metric:Eq}\\
C_{(2)}&=&\cos\theta\,\omega_{(2)}\,,\\
B^{(\rm B)}_{(2)}&=&\tan\theta(h_\theta^{-1}-1)dx\wedge dy\,,\\
\e^{2\phi_{(\rm B)}}&=&h_\theta^{-1}\,,\\
C_{(4)}&=&\omega_{(4)}\pm\sin\theta h^{-1}\Omega(\Xsp)\,,
} 
where $\Omega(\Xsp)$ is defined by \eqref{D3D5:volume-x:Eq}, and 
 $\omega_{(2)}$ and $\omega_{(4)}$ satisfy the equation 
~(\ref{D3D5:omega:Eq}). 
Finally we obtain the solution ~(\ref{D3D5:metric:Eq}) 
and (\ref{D3D5:fields:Eq}) which is derived from the dynamical D4-brane 
solution via T-duality. 

\section{The D$(p-2)$-D$p$ brane system}
\label{sec:Dp}

It is easy to obtain the dynamical solutions for other brane systems. 
Following the same procedure as the case of the D3-D5 brane, 
we can generalize the solution found in the previous section for the 
D$(p-2)$-D$p$ brane system, where $2\le p \le 6$, as follows. 

The ten-dimensional metric is written by 
\Eqr{
ds^2&=&h^{(p-1)/8}(x, z)h_\theta^{1/4}(x, z)\left[
h^{-1}(x, z)q_{\mu\nu}(\Xsp)dx^{\mu}dx^{\nu}\right.\nn\\
&&\left.+h_\theta^{-1}(x, z)\gamma_{ij}(\Ysp)dy^idy^j
+u_{ab}(\Zsp)dz^adz^b\right]\,,
 \label{s:metric:Eq}
}
where $q_{\mu\nu}(x)$, $\gamma_{ij}(y)$, and $u_{ab}(z)$ 
are the metrics for 
$(p-1)$-dimensional spacetime $\Xsp$,
the two-dimensional space $\Ysp$, and 
$(9-p)$-dimensional space $\Zsp$, respectively.
This metric form (\ref{s:metric:Eq}) is
a straightforward generalization of a bound state of a
static D$(p-2)$-D$p$ brane system with a dilaton coupling 
\cite{Cai:1999aw, Cai:2000hn, Cai:2000yk}. 
The function $h_\theta$ is given by Eq.~(\ref{D3D5:h:Eq}). 

The ten-dimensional metric \eqref{s:metric:Eq} must satisfy 
\Eqrsubl{s:Einstein equations:Eq}{
&&R_{\mu\nu}(\Xsp)=0\,,~~~R_{ij}(\Ysp)=0\,,~~~R_{ab}(\Zsp)=0\,,
   \label{s:Ricci}\\ 
&&h(x, z)=h_0(x)+h_1(z)\,;~~
D_{\mu}D_{\nu}h_0=0\,,~~\lap_{\Zsp}h_1=0\,,
  \label{s:warp:Eq}
 } 
where $D_{\mu}$ is the covariant derivative with respect to the metric 
$q_{\mu\nu}$, and $R_{\mu\nu}(\Xsp)$, 
$R_{ij}(\Ysp)$, $R_{ab}(\Zsp)$ are the Ricci tensors of the metric 
$q_{\mu\nu}$, $\gamma_{ij}$, $u_{ab}$, respectively. 

Let us consider the case $q_{\mu\nu}=\eta_{\mu\nu}$ and $u_{ab}=\delta_{ab}$
 in more detail, where $\eta_{\mu\nu}$ is the $(p-1)$-dimensional 
 Minkowski metric, and $\delta_{ab}$ are the $(9-p)$-dimensional flat 
 Euclidean metric. 
The general solution of \eqref{s:warp:Eq} is given by
\Eqr{ 
h(x, z)=c_{\mu} x^{\mu}+\tilde{c}+\sum_l\frac{M_l}{|z^a-z^a_l|^{7-p}}\,,
 \label{D(p-2)Dp:h:Eq}
}
for $p\neq 7$,
where $c_{\mu}$, $\tilde{c}$, $M_l$ and $z_l$ are 
integration constants. In the case of $p=7$,
we have
\Eqr{ 
h(x, z)=c_{\mu} x^{\mu}+\tilde{c}+\sum_l M_l\ln |z^a-z^a_l|
\,.
 \label{D5D7:h:Eq}
}
If $c_0\neq 0$, the solution  depends on time $t$.

Note that if we smear out some dimensions (e.g. $d_\Zsp \,(<9-p)$- dimensions)
 in $\Zsp$ space, the solution of \eqref{s:warp:Eq} is given by
\Eqr{ 
h(x, z)=c_{\mu} x^{\mu}+\tilde{c}+\sum_l\frac{M_l}{|z^a-z^a_l|^{7-(p+q)}}\,,
 \label{D(p-2)Dp_smear_q:h:Eq}
}
for $p+q\neq 7$, and 
\Eqr{ 
h(x, z)=c_{\mu} x^{\mu}+\tilde{c}+\sum_l M_l\ln |z^a-z^a_l|
\,.
 \label{D(p-2)Dp_smear_(7-p):h:Eq}
}
for $p+q=7$.

The dilaton field $\phi$ and
the gauge field except for $C_{(4)}$ are given by 
\Eqrsubl{s:ansatz for fields:Eq}{
&&\e^{\phi}=h^{(5-p)/4}h_\theta^{1/2}\,,
  \label{s:ansatz for scalar:Eq}\\
&&B_{(2)}=\tan\theta \left(h_\theta^{-1}-1\right)\wedge\Omega(\Ysp)\,,
  \label{s:ansatz for b:Eq}\\
&&C_{(p-1)}=\pm\sin\theta \left(h^{-1}-1\right)\wedge\Omega(\Xsp)\,,
  \label{s:ansatz for p:Eq}\\
&&C_{(p+1)}=\pm \cos\theta \left(h_\theta^{-1}-1\right)\wedge\Omega(\Xsp)
\wedge\Omega(\Ysp)\,,
  \label{s:ansatz for p+2:Eq}
}
where $C_{(p-1)}$ and $C_{(p-1)}$ are gauge potentials for 
electrically charged D$(p-2)$-brane and D$p$-brane, respectively. 
$\Omega(\Xsp)$ and $\Omega(\Ysp)$ denote the volume $(p-1)$-form
and 2-form: 
\Eqrsubl{s:volume:Eq}{
\Omega(\Xsp)&=&\sqrt{-q}\,dx^0\wedge dx^1\wedge \cdots \wedge dx^{p-2}\,,
     \label{s:volume x:Eq}\\
\Omega(\Ysp)&=&\sqrt{\gamma}\,dy^1\wedge dy^2\,,
     \label{s:volume y:Eq}
}
where $q$ and $\gamma$ are the determinants of the metric $q_{\mu\nu}$
and $\gamma_{ij}$, respectively.

This type of solution is also obtained by the procedure 
of delocalization, rotation and T duality with respect to 
more than one of the transverse coordinates of the original D-brane 
solutions. For example, we consider the D0-D2-brane with D4-brane. 
The dynamical solution can be obtained by the same procedure of 
the delocalization and rotation on a D2-brane. 
Let us single out two orthogonal planes $(y^1, y^2)$ and $(v^1, v^2)$.
If we apply the procedure of the delocalization and rotation 
on a D2-brane with respect to the $(y^1, y^2)$ plane, 
followed by T-duality map, 
we obtain the time-dependent solution for the D1-D3 brane
with an rotation angle $\theta$. 
After repeating the same procedure of the delocalization and rotation 
on a D3-brane with respect to the $(v^1, v^2)$ plane, rotating by an
angle $\psi$ 
to $(v^1, v^2)$, followed by T-duality map ~(\ref{D4:duality:Eq}), 
we can construct the solution of D0-, D2 and D4-brane system 
\cite{Breckenridge:1996tt}. We summarize this solution 
in Appendix \ref{sec:D0D2D4}.

The dynamical solutions 
certainly have many attractive properties. 
In the case of intersecting branes in supergravity, 
the field equations normally indicate that time dependent solutions 
can be found if only one harmonic function in the metric 
depends on time \cite{Maeda:2009zi, Minamitsuji:2010kb}. 
If the particular relation between dilaton couplings 
to antisymmetric tensor field strengths is satisfied, one can find 
the solutions where all harmonic functions depend on time. 
However these solutions are not apparently related to the
classical solutions of string theory \cite{Minamitsuji:2010uz}
because we have to introduce a cosmological constant.
In the present solution, two  
functions $h$ and $h_\theta$ can depend on both time and spatial
coordinates of the transverse space $\Zsp$.  
Although it is an intersecting brane solution in supergravity 
with two harmonic functions,
it turns out that there is an appropriate
 relation  (\ref{D3D5:f:Eq}) between $h$ and $h_\theta$.
It is because our solution is obtained by T-dualization
from the solution with one time-dependent brane.

Applying our dynamical solutions, we shall discuss two important cases
in the following sections: 
One is cosmology and the other is a collision of branes.

\section{Cosmological solutions}
\label{sec:cosmology}

In order to discuss cosmology by our new solution, 
we first specify which dimensions correspond to our three space.
Since our universe is isotropic and homogeneous,
three space dimensions of the universe
can be a part of a uniform brane. So if $p\geq 5$, we find 
a three-dimensional isotropic and homogeneous space in the $\Xsp$ spacetime,
which describes our three-space.
After compactifying some dimensions $(p-2)$ in $\Xsp$ space, 
whole $\Ysp$ space,
and some smeared dimensions $d_\Zsp$ in  $\Zsp$ space, 
we can regard three-dimensional part of the D$(p-2)$ branes as
our universe, which is localized in the rest uncompactified extra dimensions 
$(9-p-d_\Zsp)$ of $\Zsp$ space.
One typical example is the D3-D5 brane system.
The D3 brane in $(4-d_\Zsp)$-dimensional subspace of $\Zsp$ can be
our universe (see the U1 model of D3-D5 brane system  in Table  II).
The uniform $(2+d_\Zsp)$ dimensions are compactified.


\begin{table}[h]
\begin{center}
{\scriptsize
\begin{tabular}{|c||c|c|c|c|c|c|c|c|c|c||c|c|}
\hline
Branes&0&1&2&3&4&5&6&7&8&9&  our space $\tilde \Msp$ 
&power exponent $\lambda(\tilde{\Msp})$ \\
\hline
D0 & $\circ$ &{\scriptsize $\oplus$}&{\scriptsize $\oplus$}&&$\star$&
$\star$&$\star$&
$\rule{3pt}{3pt}$&$\rule{3pt}{3pt}$&$\rule{3pt}{3pt}$&    &
\\
\cline{2-11}
D2 & $\circ$ &{\tiny $\copyright$}& {\tiny $\copyright$} &&$\star$&$\star$&
$\star$&$\rule{3pt}{3pt}$&$\rule{3pt}{3pt}$&$\rule{3pt}{3pt}$&
$\tilde{\Zsp}$(type U2)  &
$\lambda(\tilde{\Zsp})=\displaystyle{\frac{3d_\Zsp-4}{3(d_\Zsp+4)}}$
\\
\cline{2-11}
$X^A$ & $t$ & $y^1$ & $y^2$ & $z^1$ & $z^2$ & $z^3$ & $z^4$
& $z^5$ & $z^6$ & $z^7$ & & ~$(0\leq d_\Zsp\leq 3)$
\\
\hline
\hline
D1 & $\circ$ & {\tiny $\copyright$} &{\scriptsize $\oplus$}&
{\scriptsize $\oplus$}&&$\star$
&$\star$&$\rule{3pt}{3pt}$&$\rule{3pt}{3pt}$&$\rule{3pt}{3pt}$&  & 
\\
\cline{2-11}
D3 & $\circ$ &{\tiny $\copyright$}& {\tiny $\copyright$} & {\tiny $\copyright$}
&&$\star$&$\star$&$\rule{3pt}{3pt}$&$\rule{3pt}{3pt}$&$\rule{3pt}{3pt}$& 
$\tilde{\Zsp}$(type U2)  &
$\displaystyle{\lambda(\tilde{\Zsp})=\frac{d_\Zsp-1}{d_\Zsp+3}}$
\\
\cline{2-11}
$X^A$ & $t$ & $x^1$     
& $y^1$ & $y^2$ & $z^1$ &  $z^2$ & $z^3$& $z^4$ & $z^5$
& $z^6$&  & $(0\leq d_\Zsp\leq 2)$
\\
\hline
\hline
D2 &  $\huge{\circ}$ & {\tiny $\copyright$} & {\tiny $\copyright$} &
{\scriptsize $\oplus$}&
{\scriptsize $\oplus$}&&$\star$&
$\rule{3pt}{3pt}$&$\rule{3pt}{3pt}$&$\rule{3pt}{3pt}$&  &
\\
\cline{2-11}
D4 & $\circ$ &{\tiny $\copyright$}& {\tiny $\copyright$}& {\tiny $\copyright$}
 & {\tiny $\copyright$} &&
$\star$&$\rule{3pt}{3pt}$&$\rule{3pt}{3pt}$&$\rule{3pt}{3pt}$  & 
$\tilde{\Zsp}$(type U2)  &
$\displaystyle{\lambda(\tilde{\Zsp})=\frac{5d_\Zsp-2}{5d_\Zsp+14}}$
\\
\cline{2-11}
$X^A$ & $t$ & $x^1$ & $x^2$    
& $y^1$ & $y^2$ & $z^1$ &  $z^2$ & $z^3$& $z^4$ & $z^5$&
 & $(0\leq d_\Zsp\leq 1)$
\\
\hline
\hline
D3 & $\circ$ &{\tiny $\copyright$}& {\tiny $\copyright$}& 
{\tiny $\copyright$} &{\scriptsize $\oplus$}&{\scriptsize $\oplus$}&&
$\rule{3pt}{3pt}$&
$\rule{3pt}{3pt}$&$\rule{3pt}{3pt}$&
 &
\\
\cline{2-11}
D5 & $\circ$ &{\tiny $\copyright$}& {\tiny $\copyright$} &
 {\tiny $\copyright$}& {\tiny $\copyright$} & {\tiny $\copyright$} &&
$\rule{3pt}{3pt}$&$\rule{3pt}{3pt}$&$\rule{3pt}{3pt}$& 
$\tilde{\Zsp}$(type U2)  &
$\displaystyle{\lambda(\tilde{\Zsp})=\frac{1}{9}}$
\\
\cline{2-11}
$X^A$ & $t$ & $x^1$ & $x^2$ & $x^3$   
& $y^1$ & $y^2$ & $z^1$ &  $z^2$ & $z^3$& $z^4$ &   & 
\\
\hline
\hline
D3 & $\circ$ & $\bullet$&  $\bullet$&  $\bullet$ &
{\scriptsize $\oplus$}&{\scriptsize $\oplus$}&&$\star$&
$\star$&$\star$&
 &
\\
\cline{2-11}
D5 & $\circ$ & $\bullet$ & $\bullet$&  $\bullet$ &
{\tiny $\copyright$} & 
{\tiny $\copyright$} & 
&$\star$&$\star$&$\star$&  $\tilde{\Xsp}$(type U1)  &
$\displaystyle{\lambda(\tilde{\Xsp})=\frac{3d_\Zsp-4}{3(d_\Zsp+4)}}$
\\
\cline{2-11}
$X^A$ & $t$ & $x^1$ & $x^2$ & $x^3$   
& $y^1$ & $y^2$ & $z^1$ &  $z^2$ & $z^3$& $z^4$ &  & $(0\leq d_\Zsp\leq 3)$
\\
\hline
\hline
D4 & $\circ$ & $\bullet$ & $\bullet$ & $\bullet$ & {\tiny $\copyright$} &
{\scriptsize $\oplus$}&{\scriptsize $\oplus$}&&$\star$&$\star$&
 & 
\\
\cline{2-11}
D6 & $\circ$ &$\bullet$ & $\bullet$ & $\bullet$ &{\tiny $\copyright$} & 
{\tiny $\copyright$} & {\tiny $\copyright$} &
&$\star$&$\star$&   $\tilde{\Xsp}$(type U1)  &
$\displaystyle{\lambda(\tilde{\Xsp})=\frac{7d_\Zsp-5}{7d_\Zsp+27}}$
\\
\cline{2-11}
$X^A$ & $t$ & $x^1$ & $x^2$ & $x^3$ & $x^4$  
& $y^1$ & $y^2$ & $z^1$ &  $z^2$ & $z^3$&  & $(0\leq d_\Zsp\leq 2)$
\\
\hline
\hline
D5 & $\circ$ & $\bullet$ & $\bullet$ & $\bullet$ & {\tiny $\copyright$} & 
{\tiny $\copyright$} &{\scriptsize $\oplus$}&{\scriptsize $\oplus$}&&$\star$& & \\
\cline{2-11}
D7 & $\circ$ &$\bullet$ & $\bullet$ & $\bullet$ & {\tiny $\copyright$}
& {\tiny $\copyright$} & {\tiny $\copyright$}&
 {\tiny $\copyright$} && $\star$ &  $\tilde{\Xsp}$(type U1)   &
$\displaystyle{\lambda(\tilde{\Xsp})=\frac{d_\Zsp}{d_\Zsp+4}}$
\\
\cline{2-11}
$X^A$ & $t$ & $x^1$ & $x^2$ & $x^3$ & $x^4$ & $x^5$ 
& $y^1$ & $y^2$ & $z^1$ &  $z^2$ & & $(
0\leq d_\Zsp\leq 1)$
\\
\hline
\hline
D6 & $\circ$ & $\bullet$ & $\bullet$ & $\bullet$ & {\tiny $\copyright$} & 
{\tiny $\copyright$} & {\tiny $\copyright$} &{\scriptsize $\oplus$}&
{\scriptsize $\oplus$}&&&
\\
\cline{2-11}
D8 & $\circ$ &$\bullet$ & $\bullet$ & $\bullet$ & {\tiny $\copyright$} & 
{\tiny $\copyright$} & {\tiny $\copyright$} &
{\tiny $\copyright$} & {\tiny $\copyright$} &&  $\tilde{\Xsp}$(type U1)   &
$\displaystyle{\lambda(\tilde{\Xsp})=\frac{7}{39}}$
\\
\cline{2-11}
$X^A$ & $~t~$ & $x^1$ & $x^2$ & $x^3$ & $x^4$ & $x^5$ & $x^6$
& $y^1$ & $y^2$ & $z^1$ & &  
\\
\hline
\end{tabular}
}
\end{center}
\begin{flushleft}
\baselineskip 14pt
{\small 
TABLE II
~
Brane configuration for the D$(p-2)$-D$p$ brane system and 
construction of our universe.
Depending on which dimensions are  compactified,
there are two possibilities as the candidate for our universe.
One is given by 
U1 and the other is  U2.
{\footnotesize $\copyright$} denotes
 that the corresponding brane dimension is compactified,
while $\oplus$ denotes
 that the corresponding empty dimension is smeared and compactified. 
$\star$ means that the corresponding empty dimension can be
 either smeared and compactified or vacuum as it is. 
The compactified dimension in $\Zsp$ space (the number of $\star$'s)
is $d_\Zsp$. Our three space is given by either $\bullet$ (U1) or
${\rule[.1em]{4pt}{4pt}}$ (U2).
We also show the power exponent of our three-dimensional universe 
for D$(p-2)$-D$p$ brane system.
}
\end{flushleft}
\label{table_1}
\end{table}

The other possibility is the case that we live in the three-dimensional
uniformly smeared subspace of the transverse space $\Zsp$, 
with compactification of the rest smeared subspace in $\Zsp$-space and
of the $p$-dimensional space 
in the $\Xsp\otimes\Ysp$ spacetime.
It is possible for the case of $p\leq 5$.
For example, in the case of  D3-D5 brane system,
the $\Zsp$ space is four-dimensional. 
So we smear out the three  directions $z^2,z^3$ and $z^4$.
Then we can put our universe at $z^1=z_0^1$.
As a result, we find our universe is described by the coordinates 
$(t,z^2,z^3,z^4)$ with $5$-dimensional compactified space
 (see the U2 model of D3-D5 brane system in Table  II).

Although both solutions are exact,
our 3-space in the latter case is an aggregation of the smeared branes and 
it is unclear how we can recover four-dimensional gravity in our universe,
except for the conventional Kaluza-Klein realization of 4D gravity 
by compactification.
On the other hand, since our 3-space in the former case is 
a part of branes, we may invoke a brane world scenario
\cite{brane_world0,brane_world1,brane_world2,brane_world3}.

Here we show how to construct our universe from D$(p-2)$-D$p$ brane system in the former case.
The ten-dimensional metric in the D$(p-2)$-D$p$ brane system
is expressed as
\Eq{
ds^2=-h^{(p-9)/8}h_\theta^{1/4}\left[-dt^2+ds^2(\tilde{\Xsp})\right]
+h^{(p-1)/8}h_\theta^{-3/4}ds^2(\Ysp)+h^{(p-1)/8}h_\theta^{1/4}ds^2(\Zsp)\,,
   \label{c:metric:Eq}
}
where 
\Eqrsubl{c:metric1:Eq}{
ds^2(\tilde{\Xsp})&\equiv&
\delta_{pq}(\tilde{\Xsp})dx^{p}dx^{q}\,,\\
ds^2(\Ysp)&\equiv&\gamma_{ij}(\Ysp)dy^idy^j\,,\\
ds^2(\Zsp)&\equiv&u_{ab}(\Zsp)dz^adz^b\,.
 }
Here $\tilde{\Xsp}$ is the $(p-2)$-dimensional Euclidean space, which 
coordinates are described by $x^p$
and the metric is given by $\delta_{PQ}(\tilde{\Xsp})$. 
$h$ and $h_\theta$ are given by 
$h_\theta=1+\cos^2\theta(h-1)$ and $h=c_0 t+h_1(z)$,
 where $c_0$ is a constant. 
Here, in order to discuss an isotropic and homogeneous universe,
we assume that $c_i=0$ in (\ref{D(p-2)Dp_smear_q:h:Eq}).

Now we  compactify $d\,(\equiv d_\Xsp+d_\Ysp+d_\Zsp)$ dimensional space
 to find our universe,
where $d_\Xsp$, $d_\Ysp$, and $d_\Zsp$ denote the compactified dimensions with
respect to the $\tilde{\Xsp}$, $\Ysp$, and $\Zsp$ spaces. 
$d_\Xsp=(p-2)-3=p-5$ is the compactified dimensions of $(p-2)$-branes.
All dimensions of $\Ysp$ must be compactified, i.e., $d_\Ysp=2$.
We also consider the possibility of smearing out
some dimensions ($d_\Zsp$) of $\Zsp$ space, which are compactied.
Since the function $h_1$ depends on the transverse directions, 
$d_\Zsp$ should satisfy $d_\Zsp<9-p$\,. 
The metric (\ref{c:metric:Eq}) is then described by
\Eq{
ds^2=h^Bh_\theta^C ds_D^2(\bar{\Msp})
+h^{p-1\over 8} h_\theta^{1\over 4} u_{\bar{a}\bar{b}}(\bar{\Zsp})
dz^{\bar{a}}dz^{\bar{b}}+ds^2(\Nsp')\,,
   \label{c:metric2:Eq}
}
where the exponents $B$ and $C$ are defined by
\Eqr{
B&=&-{d(p-1)\over 8(D-2)}+{d_\Xsp\over D-2}
\nn\\
C&=&-{d\over 4(D-2)}+{d_\Ysp\over D-2}
\label{exponent:Eq}
\,.
}
We have used a bar for the variables in uncompactified space
$ds_D^2(\bar{\Msp})$ is the $D$-dimensional metric in the Einstein frame, 
which is  given by
\Eq{
ds_D^2(\bar{\Msp})=h^\alpha h_\theta^\beta \left[-dt^2+\delta_{\bar{p}\bar{q}}
(\bar{\Xsp})dx^{\bar{p}}dx^{\bar{q}}\right]
\,,
}
with 
\Eqr{
D&=&p-1-d_\Xsp
\nn\\
\alpha&=&{1\over 8(D-2)}\left[(D-2)(p-9)+d(p-1)-8d_\Xsp\right]\,, 
\nn\\
\beta&=&{1\over 4(D-2)}\left[D-2+d-4d_\Ysp\right]
\,,
}
and $\bar{\Zsp}$ is the uncompactified transverse space with the coordinates
$z^{\bar{a}}$.
While 
a prime is used for the variables in the compactified space, i.e., 
$ds^2(\Nsp')$ is the metric of compactified dimensions $\Nsp'$, 
which is given by
\Eq{
ds^2(\Nsp')=h^{p-9\over 8} h_\theta^{1\over 4} \delta_{p'q'}
(\tilde{\Xsp}')dx^{p'}dx^{q'}
+h^{p-1\over 8} h_\theta^{-{3\over 4}} \delta_{ij}
dy^{i}dy^{j}
+h^{p-1\over 8} h_\theta^{1\over 4} \delta_{a'b'}
dz^{a'}dz^{b'}
\,.
}

Our universe is described by $ds_D^2(\bar{\Msp})$ on a D$(p-2)$-brane.
The dimension of the uncompactified transverse  space $\bar{\Zsp}$ is
$9-p-d_\Zsp$. Although the warp factor $h$ diverges on the brane
unless $9-p-d_\Zsp=1$, 
we expect that it will be regularized by a stringy effect.
Hence  we shall evaluate $h$ at $z_0^{~\bar{a}}$ near the brane, i.e.,
 $h_1(z_0^{~\bar{a}})=\xi$ finite constant.
For the case of $9-p-d_\Zsp=1$ (dim ($\bar{\Zsp}$)=1),
 the warp factor is finite on the brane without a stringy effect.

As a result, we evaluate the metric of our universe as
\Eq{
ds_D^2=(\cos^{2}\theta)^\beta h^\alpha (h+\tan^2\theta)^\beta\left(
-dt^2+d\,\bar{\vect{x}}^2\right)
\,,
}
where 
$h=c_0 t+h_1(z_0^{~\bar{a}})$.
When $c_0>0$, in the limit of $t\rightarrow \infty$, introducing 
a cosmic time $\tau$  by
\Eq{
d\tau=(\cos \theta)^\beta c_0^{{1\over 2}(\alpha+\beta)}
t^{{1\over 2}(\alpha+\beta)}dt
}
or
\Eq{
\tau\propto 
t^{{1\over 2}(\alpha+\beta+2)}
\,,
}
we find the scale factor of our universe as
\Eq{
a_E(\tau)\approx 
(\cos \theta)^\beta c_0^{{1\over 2}(\alpha+\beta)} t^{{1\over 2}(\alpha+\beta)}
\propto \tau^\lambda
\,
}
where 
\Eqr{
\lambda&=&{\alpha+\beta\over \alpha+\beta+2}
\nn\\
&=& {(p+1)d_\Zsp+(p-1)(p-7)\over (p+1)d_\Zsp+p^2+8p-41-16d_\Xsp}
\\
&=&{(p+1)d_\Zsp+(p-1)(p-7)\over (p+1)d_\Zsp+p^2-8p+39}
\,.
}
In the last equality, we set $D=4$.

In the case of the D3-D5 brane system, we find
\Eq{
\lambda={3d_\Zsp-4\over 3(d_\Zsp+4)}
\,.
}
For $d_\Zsp= 0, 1, 2$, 
and 3, we obtain $\lambda=-1/3, -1/15, 1/9$, 
and $5/21$, respectively.

We can perform the similar analysis to construct the type U2 universe 
from D$(p-2)$-D$p$ brane system.
Our three space is given by three-dimensional subspace
 $(z^{7-p}, z^{8-p}, z^{9-p})$ (marked by $\rule{5pt}{5pt}$ in Table II)
in $\Zsp$ space, which is smeared to obtain our homogeneous and isotropic 
universe. The power exponent $\lambda$  of the scale factor is given by
\Eq{
\lambda={(p+1)d_\Zsp +p(p-5)+2 \over (p+1)d_\Zsp +p(p-5)+18 }
\,.
}

In Table II, we summarize the possible cosmological
solutions (both U1 and U2) derived 
 from D$(p-2)$-D$p$ brane systems and the power exponent $\lambda$  
of the scale 
factors of our universe. The maximum value of $\lambda$ is 5/21 
in the case of D3-D5 brane system for the type U1 universe, 
and in the case of D0-D2 brane system for the type U2 universe.
Although we find the exact time-dependent brane solution,
the power exponent of our scale factor may be too small to explain
our expanding universe.
Furthermore, in order to discuss an inflationary scenario in an interacting 
brane system, one may need additional 
ingredients such as a brane-antibrane interaction,
which is beyond our present approach.

\section{Collision of branes (or universes)} 
\label{sec:collision}
Next, we apply our dynamical solutions to 
a collision of $N$ brane systems. 
If we construct a universe from each brane system
by compactification as shown in \S.\ref{sec:cosmology},
our solution describes collision of $N$  universes.

As the case of cosmology, $h$ is assumed to be
\Eq{
h(x, z)=c_0 t +\tilde c+h_1(z)\,,
 \label{co:h2:Eq}
}
where $c_0$ and $\tilde c$ are constant parameters, and
the harmonic function $h_1$ is given by
\Eqrsubl{co:L:Eq}{
h_1(z)&=&\sum_{l=1}^{N}\frac{M_l}{|\,\vect{z}-\,\vect{z}_l|^{7-(p+d_\Zsp)}}\,,
~~~~~~{\rm for}~~p\neq 7-d_\Zsp\,,
\label{harmonics_neq4}
\\
h_1(z)&=&\sum_{l=1}^{N}M_l\, \ln |\,\vect{z}-\,\vect{z}_l|\,,
~~~~{\rm for}~~p= 7-d_\Zsp
\,,
\label{harmonics_4}
}
where 
$M_l~ (l=1\,,\cdots\,, N)$ are mass constants of D$p$-branes
located at $\,\vect{z}_l$ and 
\Eq{
|\,\vect{z}-\,\vect{z}_l|
=\sqrt{\left(z^1-z^1_l\right)^2+\left(z^2-z^2_l\right)^2+\cdots+
\left(z^{9-(p+d_\Zsp)}-z^{9-(p+d_\Zsp)}_l\right)^2}\,,
}
 because $h_1$ is the harmonic function on the
$[9-(p+d_\Zsp)]$-dimensional Euclidean subspace in $\Zsp$.
The metric, dilaton, and gauge field strength of the solution are
 given by Eqs.~(\ref{s:metric:Eq}), and
(\ref{s:ansatz for fields:Eq}), respectively.
We see that D$(7-d_\Zsp)$-brane case is critical.
For D$(8-d_\Zsp)$-brane, the function $h_1$ is given by 
the sum of linear functions of $z$.
The difference in the transverse dimensions
causes significant difference in the behaviors of the gravitational field
strengths in the transverse space, and the possibility of brane collisions.

Note that the ten-dimensional metric (\ref{s:metric:Eq}) is regular 
if and only if  
$h>0$ and $h_\theta>0$, but the spacetime 
shows curvature singularities at $h=0$ or at $h_\theta=0$. 
So the regular ten-dimensional spacetime
 is restricted to the region of $h>0$ and $h_\theta>0$, which is 
bounded by curvature singularities.

The solution (\ref{c:metric:Eq}) with $N$ D$(p-2)$-D$p$-branes 
takes the form
\Eqr{
ds^2&=&\cos^{1\over 2}\theta\left[c_0 t+h_1(z)\right]^{(p-1)/8}
\left[\tan^2\theta+c_0t+h_1(z)\right]^{1/4}\left[
{1\over c_0t+h_1(z)}
\eta_{\mu\nu}dx^{\mu}dx^{\nu}\right.\nn\\
&&\left.+{1\over \cos^2\theta\left(\tan^2\theta+c_0t+h_1(z)
\right)}\delta_{ij}dy^idy^j+\delta_{ab}dz^adz^b\right]\,,
   \label{co:sufrace:Eq}
}
where we set $h_0=c_0 t$
and the function $h_1$ is defined in (\ref{co:L:Eq}).
The behavior of the harmonic function $h_1$ is classified into two
classes depending on the dimensions of the D-brane $p$, that is, 
$p\le (6-d_\Zsp)$ and $p= (8-d_\Zsp)$, 
which we will discuss below separately.
For the D$(7-d_\Zsp)$-brane, the harmonic function $h_0$ diverges
both at infinity and near D$(7-d_\Zsp)$-branes.
In particular, because $h_1\rightarrow -\infty$,
there is no regular spacetime region near branes.
Hence, such solutions are not physically relevant.

\subsection{Collision of the D$(p-2)$-D$p$ brane ($p\le (6-d_\Zsp)$)}
\label{p<pcr}

In the limit of $\,\vect{z}\rightarrow \,\vect{z}_l$
(near branes),
the harmonic function $h_1$ becomes dominant.
Hence, we find a static  structure of D$(p-2)$-D$p$ brane system.
On the other hand, in the far region from branes, i.e.,
in the limit of $|\,\vect{z}|\rightarrow \infty$,
the function $h$ depends only on time $t$ because
$h_1$ vanishes. The metric is thus given by
\Eqr{
ds^2&=&\left(c_0t\right)^{(p-1)\over 8}\cos^{1\over 2}\theta
\left(c_0t+\tan^2\theta\right)^{1\over 4}\left[
\left(c_0t\right)^{-1}\eta_{\mu\nu}dx^{\mu}dx^{\nu}\right.\nn\\
&&\left.+{1\over \cos^2\theta\left(c_0t+\tan^2\theta\right)}
\delta_{ij}dy^idy^j+\delta_{ab}dz^adz^b
+\delta_{a'b'}dy^{a'}dy^{b'}+\delta_{\bar a\bar b}dz^{\bar a}dz^{\bar b}
\right]
\,.
   \label{co:sufrace2:Eq}
}

To study more detail, we shall analyze one concrete example, in which 
two branes, 
 are located at $\,\vect{z}=(\pm L, 0,\cdots,0)$.
Since the behavior of spacetime highly depends on the signature of
a constant $c_0$, we discuss the dynamics separately.
The metric function is singular at zeros
of the function (\ref{co:h2:Eq}). Namely
the regular spacetime exists inside the domain restricted by
\Eq{
h(t, z) = c_0t+h_1(z)>0\,,~~~~~h_\theta(t, z) = \cos^2\theta
\left[\tan^2\theta
+c_0t+h_1(z)\right]>0\,,
}
where the function $h_1$ is defined in (\ref{co:L:Eq}).
The spacetime cannot be extended beyond this region, because
not only the scalar field
$\phi$ diverges but also the spacetime evolves into 
a curvature singularity.

The regular spacetime with two D$p$-branes 
($p+d_\Zsp\le 6$) ends on these singular
hypersurfaces.
Since the time dependence appears only in the form of $c_0 t$,
the solution with $c_0>0$ is the time reversal one of $c_0<0$.
Hence we will analyze the case with $c_0<0$ in what follows.

For  $t<0$, as the function $h$ is positive everywhere and  
the ten-dimensional spacetime is nonsingular. In the limit 
of $t\rightarrow -\infty$, it is asymptotically 
a time dependent uniform spacetime except for 
near branes, where the cylindrical forms of infinite 
throats exist. 
  
When $t>0$, the spatial metric is initially ($t=0$) regular everywhere and 
the spacetime has a cylindrical topology near each brane. 
As $t$ increases slightly, a singularity 
appears from a far region ($|\,\vect{z}-\,\vect{z}_l|\rightarrow\infty$). 
As $t$ increases further, the singularity cuts off more and more of the 
space. As $t$ continues to increase, the singularity eventually splits 
and surrounds each of the brane throats individually. 
The spatial surface is then composed of two isolated throats.

The metric (\ref{co:sufrace2:Eq}) implies that the transverse 
dimensions expand asymptotically as $t^{(p-1)/8}$ 
for fixed spatial coordinates ($z^a$)\,. 
However, it is observer-dependent.
As we mentioned before, it is static near branes, and
the spacetime approaches a FLRW 
universe in the far region ($|\,\vect{z}-\,\vect{z}_l|\rightarrow \infty$),
which expands in all directions isotropically.
For the period of $t<0$, the behavior of spacetime is
the time reversal of the period of $t>0$.

\begin{figure}[h]
 \begin{center}
  \includegraphics[keepaspectratio, scale=.7]{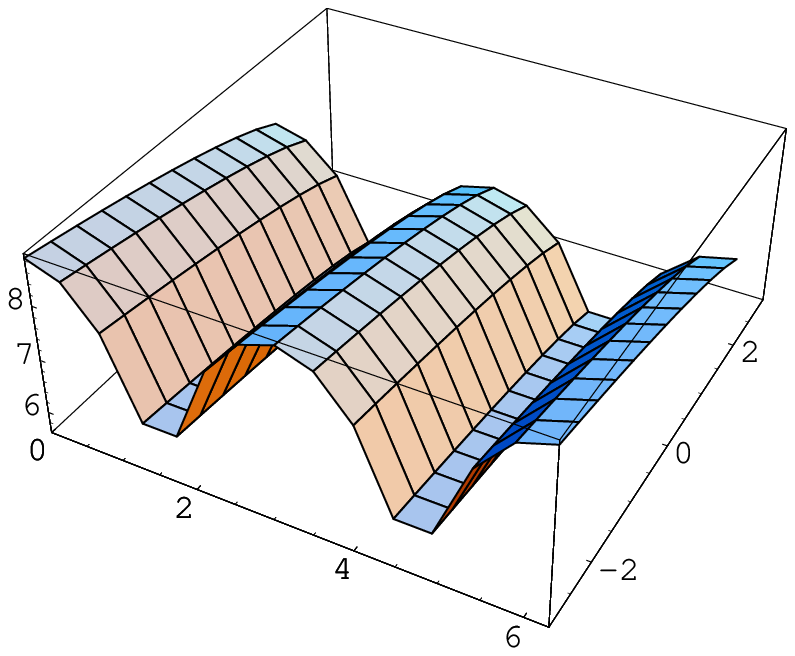}
\put(-180,85){$d(t)$}
\put(-100,0){$\theta$}
\put(0,75){$t$}
\hskip 1.5cm
\includegraphics[keepaspectratio, scale=.8]{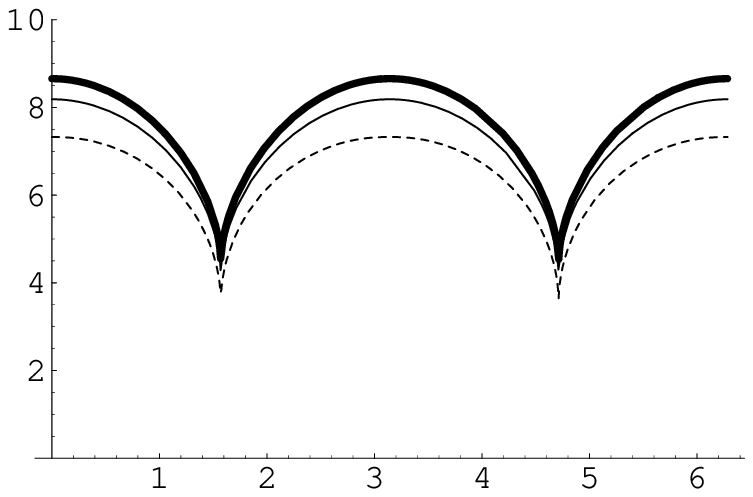}
\put(-180,125){$d(t)$}
\put(0,10){$\theta$}
\\
 (a)\hskip 8cm (b)~~~~~~~~~
  \caption{\baselineskip 14pt
(a) The proper distance between two D5-branes 
for D3-D5 brane system given in 
  \eqref{distance} is depicted (a). It decreases monotonically as time 
  increases. We set  $M_1=1$, $M_2=1$, 
  $c_0=-1$ and $L=1$. A singularity appears between two D5-branes
  at $t=2$ and the spacetime split into two isolated brane throats  
before they collide. 
(b) We also show the snapshots
 at $t=-2$ (bold),$0$ (solid), and $2$(dashed) from the top. 
 Although the distance depends sensitively 
on the angle $\theta$,
but not on time $t$.
  }
  \label{fig:D5-2}
 \end{center}
\end{figure}

\begin{figure}[h]
 \begin{center}
\includegraphics[keepaspectratio, scale=.8]{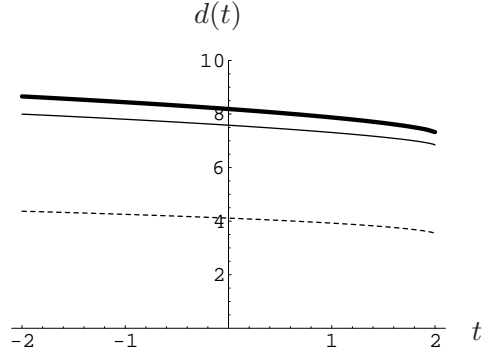}
\put(-105,125){$d(t)$}
\put(0,5){$t$}
\\
  \caption{\baselineskip 14pt
The time change of the proper distance between two D5-branes 
for D3-D5 brane system at $\theta=0$ (bold), $\pi/4$ (solid) 
and $\pi/2$ (dashed).
We choose the same parameters as Fig.\ref{fig:D5-2} ($M_1=1$, $M_2=1$, 
  $c_0=-1$ and $L=1$).  Two branes approach very slowly and a singularity 
appears at $t=2$.
  }
  \label{fig:D5-3}
 \end{center}
\end{figure}

Defining  
\Eq{
z_{\perp}=\sqrt{\left(z^2\right)^2+\left(z^3\right)^2+\cdots 
+\left(z^{9-(p+d_\Zsp)}\right)^2}\,,
}
the proper distance at $z_{\perp}=0$
between two branes is given by
\begin{eqnarray}
d(t)&=&\cos^{1\over 4}\theta\int_{-L}^L dz^1 
\left[c_0t+{M\over |z^1+L|^{7-p}}
+{M\over |z^1-L|^{7-p}}\right]^{(p-1)/16}\nn\\
&&~~~~~~~~~~~~~~~~~
\times\left[\tan^2\theta+c_0t+{M\over |z^1+L|^{7-p}}
+{M\over |z^1-L|^{7-p}}\right]^{1/8}
\,,
\label{distance}
\end{eqnarray}
which is a monotonically increasing function of $t$.
In Figs. \ref{fig:D5-2} and \ref{fig:D5-3}, we  show $d(t)$  
for the case of D3-D5 brane system.
We choose $c_0=-1$, $M_1=M_2=1$ and $L=1$. 
Initially ($t<0$), all of the region of ten-dimensional space is regular
except at $|\,\vect{z}-\,\vect{z}_l|\rightarrow 0$.
 They are asymptotically 
time dependent spacetime and have the cylindrical form of an infinite 
throat near the D5-brane.   
At $t=0$, the singularity  appears 
from a far region ($|\,\vect{z}-\,\vect{z}_l|\rightarrow\infty$).
As time evolves ($t>0$), 
the singular hypersurface erodes the region with 
the large values of $|\,\vect{z}-\,\vect{z}_l|$.
As a result, only the region of near D5-branes remains regular.
A singular hypersurface eventually surrounds each D5-brane
individually at $t=2$ and then the regular regions near D5-branes splits
into two isolated throats.
However Figs. \ref{fig:D5-2} and \ref{fig:D5-3} show
 that this singularity appears 
 before the distance $d$ vanishes, i.e., 
a singularity between two branes forms before their collision.
Two branes approach very slowly, a singularity suddenly appears
at a finite distance and the spacetime splits
into two isolated brane throats. 
Hence,
we cannot discuss a brane collision in this example.

\subsection{Collision of the D$(p-2)$-D$p$ brane ($p=(8-d_\Zsp)$)}

In this case, we have one uncompactified extra dimension $z$ in $\Zsp$ space.
Since the harmonic function $h_1$ is linear in $z$, 
we find difference behavior from the case (\ref{p<pcr}).
In order to discuss the detail, we consider one concrete example, i.e.,
the D3-D5 brane system which are smeared in the three 
transverse directions as well as $\Ysp$ space
(see Table II).

Here we assume that D3-D5 brane system are smeared 
along $z^2$, $z^3$, $z^4$ directions. 
The ten-dimensional metric (\ref{c:metric1:Eq}) 
can be written as
\Eqr{
ds^2&=&\cos^{1\over 2}\theta\left(c_0t+h_1(z)\right)^{1/2}
\left(
\tan^2\theta+c_0t+h_1(z)\right)^{1/4}
\left[
{1\over c_0t+h_1(z)}\eta_{\mu\nu}dx^{\mu}dx^{\nu}
\right.
\nonumber
\\
&&
\left.
+{1\over \cos^2\theta\left(\tan^2\theta+c_0t+h_1(z)\right)}
\delta_{ij}dy^idy^j
+\delta_{ab}dz^adz^b
\right]\,,
\label{cD3:metric-a:Eq}
}
where we set $z=z^1$ and 
the harmonic function $h_1(z)$ is given by 
\Eqr{
h_1(z)=\sum_{l=1}^{N}M_l\,|\,z-\,z_l|\,.
    \label{cD3:L:Eq}
}

We analyze the D3-D5 brane system with the brane charge $M_1$ at $z=0$
and the other $M_2$ at $z=L$.
The proper distance between the two D5-branes is given by
\Eqr{
d(t)&=&\cos^{1\over 4}\theta\int^L_0 dz \left(c_0t+
 M_1|z|+M_2|z-L|\right)^{1/4}\nn\\
 &&\times\left[\tan^2\theta+c_0t+
 M_1|z|+M_2|z-L|\right]^{1/8}\,.
  \label{D5:length:Eq}
}
For $c_0<0$,  the proper distance decreases as
$t$ increases, and if $M_1\ne M_2$, a singularity appears at 
$t=t_s\equiv-[M_1|z|+M_2|z-L|]/c_0>0$ 
when the distance is still finite. 
This is just the same as the case in \S. \ref{p<pcr}.
However, if  $M_1=M_2=M$, the result changes completely.
The distance eventually vanishes at $t=t_s:=-ML/c_0$
as 
\Eqr{
d(t)\approx \sin^{1\over 4}\theta |c_0|^{1\over 4}L(t_s-t)^{1\over 4}
\propto a^{2\over 3}
\,,
}
and two branes collide completely.
A singularity appears at the same time.
Note that the scale factor $a$ of our universe behaves as $a\propto
(t_s-t)^{3/8}$ near collision. 

\begin{figure}[h]
 \begin{center}
     \includegraphics[keepaspectratio, scale=0.6]{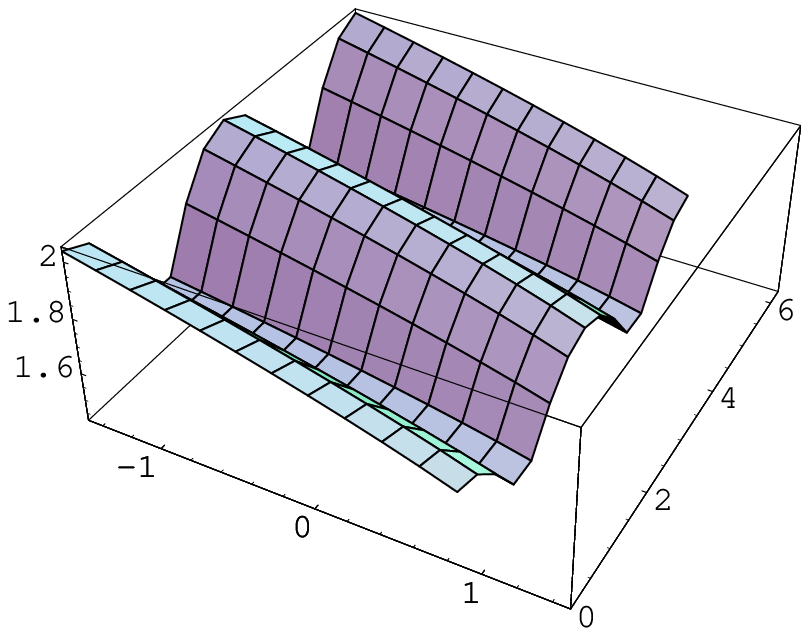}
\put(-140, 80){$d(t)$}
\put(-15, 25){$\theta$}
\put(-100,10){$t$}
\hskip 1.5cm
\includegraphics[keepaspectratio, scale=0.8]{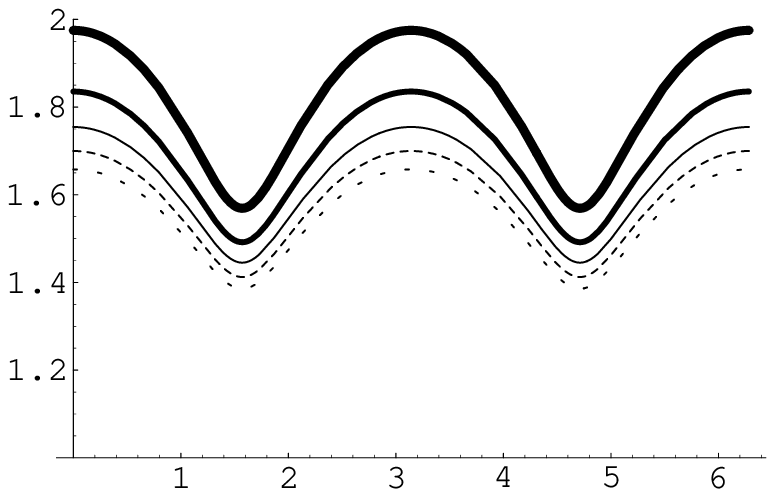}
\put(-180,125){$d(t)$}
\put(10,10){$\theta$}\\
(a) \hskip 7cm (b) ~~~~~~
  \caption{\baselineskip 14pt
(a) For the case of $M_1\ne M_2$, the proper distance given in 
  \eqref{D5:length:Eq} is depicted. We fix $c_0=-1$, $M_1=10, 
M_2=1$ and $L=1$. 
(b) We also show the snapshots at $t=-1$ (thick bold),
 $0$ (bold), $0.5$ (solid), $0.8$ (dashed), and $1$
(dotted) from the top. 
 Although the proper distance decreases as $t$ increases,
the distance 
  is still finite when 
a singularity appears  at $t=1$ on the brane located at $z=0$.}

  \label{fig:D5s1}
 \end{center}
\end{figure}

\begin{figure}[h]
 \begin{center}
\includegraphics[keepaspectratio, scale=0.6]{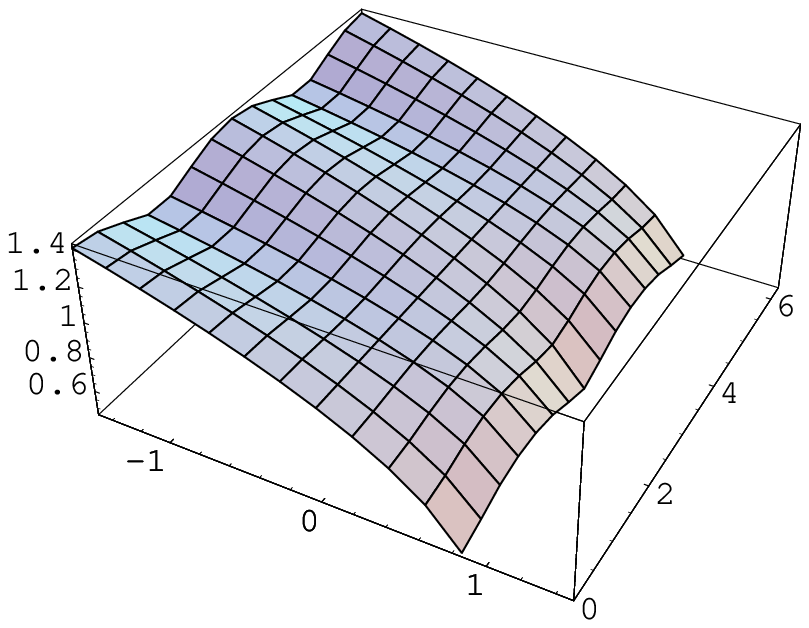}
\put(-140, 80){$d(t)$}
\put(-15, 25){$\theta$}
\put(-100,10){$t$}
\hskip 1.5cm
\includegraphics[keepaspectratio, scale=0.8]{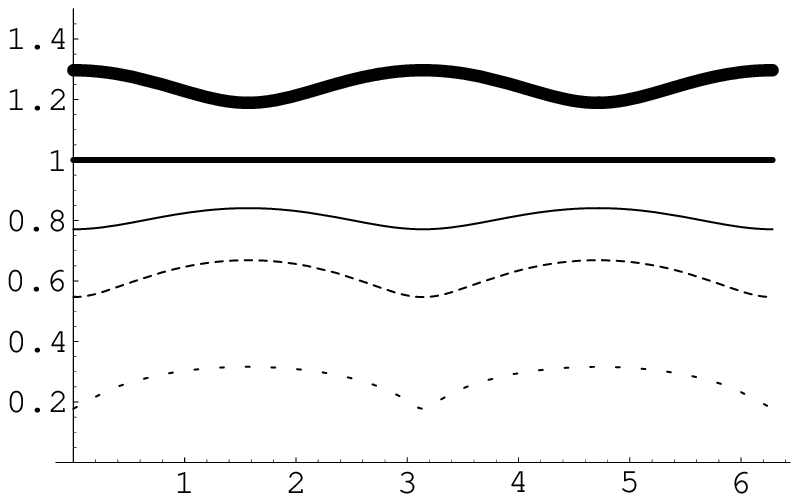}
\put(-180,125){$d(t)$}
\put(10,10){$\theta$}
\\
(a) \hskip 7cm (b) ~~~~~~
\\
  \caption{\baselineskip 14pt
(a) For the case of $M_1= M_2$, the time change of 
the proper distance given in 
  \eqref{D5:length:Eq} is depicted (a). We fix $c_0=-1$,
 $M_1=1,M_2=1$ and $L=1$.
 The proper distance decreases as $t$ increases
and it vanishes at $t=t_s>0$ when a singularity appears.
(b) We also show the snapshots at $t=-1$ (thick bold),
 $0$ (bold), $0.5$ (solid), $0.8$ (dashed), and $0.99$
(dotted) from the top.  $d$ vanishes at $t=1$. 
}
  \label{fig:D5s3-2}
 \end{center}
\end{figure}

\begin{figure}[h]
 \begin{center}
  \includegraphics[keepaspectratio, scale=0.8]{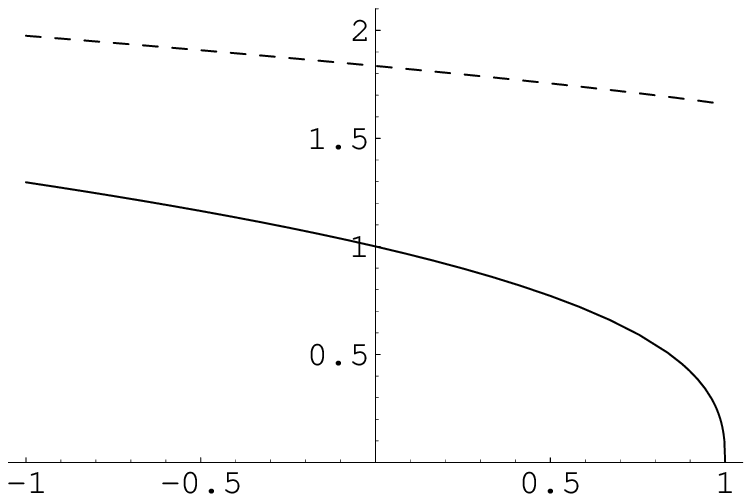}
\put(-105, 125){$d(t)$}
\put(0,10){$t$}
\hskip 1.5cm
\includegraphics[keepaspectratio, scale=0.8]{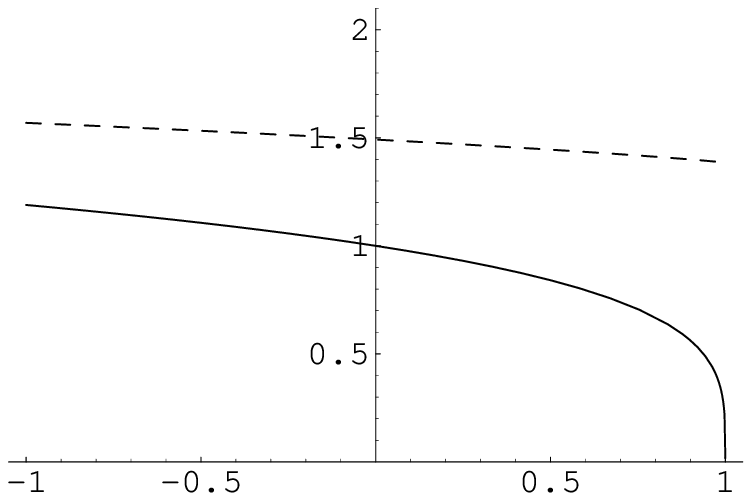}
\put(-105,125){$d(t)$}
\put(0,10){$t$}
\\
(a) \hskip 7.5cm (b) ~~~~~~
\\
  \caption{\baselineskip 14pt
The time change of 
the proper distance at $\theta=0$ (a) and
$\theta=\pi/2$ (b) for $M_1=M_2=1$ and $M_1=10,M_2=1$. We fix $c_0=-1$,
 and $L=1$.
 The proper distance rapidly vanishes near two branes collide for 
the case of $M_1=M_2=1$. 
The dashed line denotes the case of $M_1=10,M_2=1$ while 
the solid one corresponds to $M_1=M_2=1$ case.
While for the case of 
$M_1=10,M_2=1$, 
it is still finite when a singularity appears.}
  \label{fig:c9-2}
 \end{center}
\end{figure}

We show $d(t)$ integrated numerically in Fig. \ref{fig:D5s1} - 
Fig. \ref{fig:c9-2}
for the case of $c_0<0$.

In the past direction, the distance $d$ increases.
Then, for $t<0$, each brane gradually
separates as $|t|$ increases.

\subsection{Collision of brane universes in a 
lower-dimensional effective theory}

Next we consider the brane collision in the lower-dimensional 
effective theory. 
It is motivated by a brane world scenario, which is 
modeled in five dimensions after compactification
\cite{brane_world0,brane_world1,brane_world2,brane_world3}.

We compactify the $\Ysp$ space and some directions of $\Zsp$,
 where D3-D5 branes 
have been smeared along such directions.
As a result, we find  
the five-dimensional metric in the Einstein frame  as 
\Eqr{
ds^2(\bar{\Msp})&=&
\left[c_0t+h_1(z)\right]^{1/3}
\left(-dt^2+\delta_{pq}dx^pdx^q\right)
+\left[c_0t+h_1(z)\right]^{4/3}
dz^2\,,
  \label{co:metric:Eq}
}
where 
$h_1(z)$ is given by \eqref{cD3:L:Eq}\,. 
The five-dimensional metric turns out not to depend on $\theta$\,,
which makes some difference between analysis in full ten dimensions 
and that in the effective theory.

Suppose that our universe is given by D3 brane at $z=0$ and the other brane 
universe exists at $z=L$.
The metric of our universe is given by
the four-dimensional Einstein frame 
from the five-dimensional metric (\ref{co:metric:Eq}) as
\Eqr{
ds_4^2&=&
\left[c_0t+h_1(0)\right]
\left(-dt^2+\delta_{pq}dx^pdx^q\right)
\,,
  \label{co:metric2:Eq}
}
where $h_1(0)=M_2L$.
Introducing the cosmic time $\tau$ in four dimensions by
\begin{eqnarray}
{\tau_s-\tau \over \tau_0}
=\left({t_s-t \over t_0}\right)^{3\over 2}
\,,
  \label{cosmic_time}
\end{eqnarray}
where $t_0=-1/c_0$ and $\tau_0=2t_0/3$.
$t_s=-M_2L/c_0$ and an integration constant $\tau_s$ 
correspond to times when a singularity appears in 
each time coordinate.
The scale factor $a_5$ of our universe  in the
 effective 5-dimensional spacetime  is given by
\begin{eqnarray}
a_5=\left({t_s-t \over t_0}\right)^{1\over 2}
=\left({\tau_s-\tau \over \tau_0}\right)^{1\over 3}
~~~~~{\rm for}~~~\tau<\tau_s
\,.
  \label{scale_factor_IN5D}
\end{eqnarray}

The proper distance between two universes in this
 effective 5-dimensional spacetime is 
\begin{eqnarray}
d_5(t)
=\left\{
\begin{array}{ll}
\displaystyle{\left({t_s-t\over t_0}\right)^{2/3}L}&
~~~~{\rm for }~~M_1=M_2
\nonumber 
\\
\displaystyle{\frac{3}{5(M_1-M_2)}\left[\left({t_s-t\over t_0}+(M_1-M_2)L\right)^{5/3}
-\left({t_s-t\over t_0}\right)^{5/3}\right]}&
~~~~{\rm for }~~M_1\neq M_2
\\
\end{array}
\right.
  \label{D5E:length:Eq}
\end{eqnarray}

As $t$ increases,
 the proper distance $d$ decreases and it eventually 
vanishes at $t=t_s\equiv ML\tau_0(>0)$ if $M_1=M_2=M$.
When two branes approach, both universes are contracting,
and a big crunch singularity appears when two branes collide.
We find that a complete collision occurs simultaneously at $t=-ML/c_0$.
The distance vanishes as $d_5\propto a_5^{4/3}$ near collision.
 
\begin{figure}[h]
 \begin{center}
\includegraphics[keepaspectratio, scale=0.8]{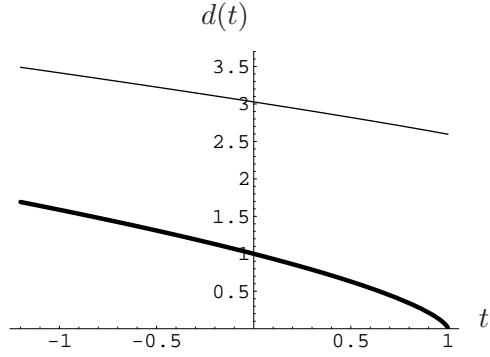}
\put(-105,125){$d(t)$}
\put(0,10){$t$}
\\
  \caption{\baselineskip 14pt
The proper distance $d_5$
is depicted. We fix $c_0=-1$ and $L=1$. 
  The proper distance decreases as $t$ increases. 
  The bold line denotes the case of $M_1=10,M_2=1$ while 
 the solid ones corresponds to $M_1=M_2=1$ case.
  If we set $M_1=M_2$, 
  it causes the complete collision at $t=t_s (=1)$ simultaneously. 
  For $M_1\ne M_2$, a singularity appears at $t=t_s$ when the distance 
  is still finite. Then, the solution cannot describe collision of 
two  branes.
}
  \label{fig:D5c1}
 \end{center}
\end{figure}

On the other hand, for the case of $M_1\ne M_2$, a singularity appears at 
at $t=t_s\equiv-[M_1|z|+M_2|z-L|]/c_0>0$, when the distance is still finite. 
We show $d(t)$ integrated numerically in Fig. \ref{fig:D5c1}.

\begin{figure}[h]
 \begin{center}
  \includegraphics[keepaspectratio, scale=0.8]{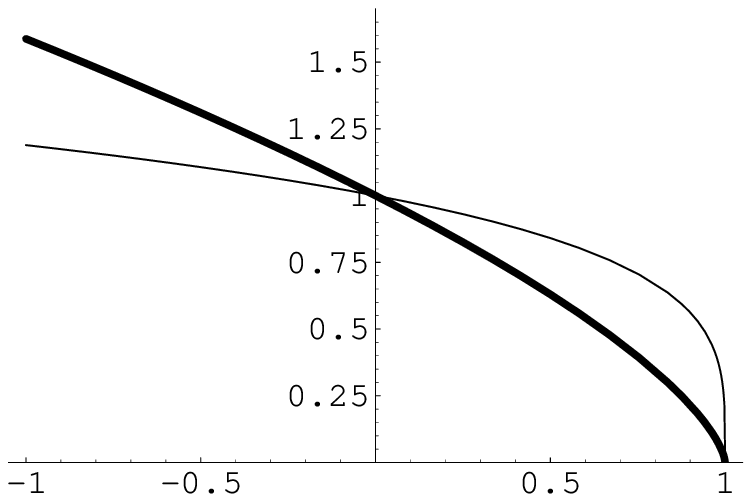}
\put(-105, 125){$d(t)$}
\put(0,10){$t$}
\hskip 1.5cm
\includegraphics[keepaspectratio, scale=0.8]{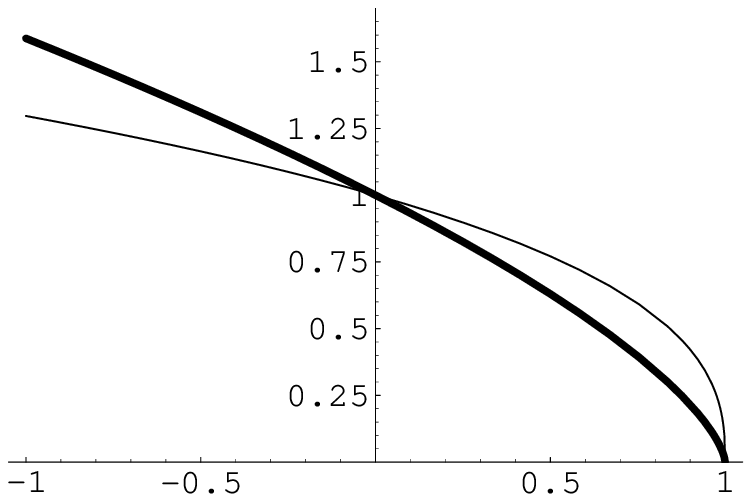}
\put(-105,125){$d(t)$}
\put(0,10){$t$}
\\
(a) $\theta=\pi/2$\hskip 6.5cm
(b) $\theta=0$\\
  \caption{\baselineskip 14pt
The proper distance given in the effective theory is compared 
with that in 10-D theory ((a) $\theta=\pi/2$
(b) $\theta=0$). The bold line denotes the proper distance in the 
effective theory while the solid ones corresponds to the proper 
distance in the 10-D theory.
}
  \label{fig:comparison}
 \end{center}
\end{figure}

We also show the comparison of the distance evaluated in the effective five 
dimensions and that in the original ten dimensions 
in Fig. \ref{fig:comparison}.
We find the behaviors are quite similar, i.e.,
two branes collide at a big crunch singularity if $M_1=M_2$,
but it is not the case for  $M_1\neq M_2$. 
However there exist
 quantitative differences, especially the approaching velocity 
of two branes in ten dimensions is much faster than that 
in the effective five dimensions when two branes collide
($d\propto (t_s-t)^{1/4}$ and $d_5\propto (t_s-t)^{2/3}$).
Hence the real collision of two branes will be much more violent
than that expected from the effective model.

\section{Conclusion and remarks}
  \label{sec:discussions}

In this paper, we have derived  the time-dependent solutions corresponding to 
the dynamical D-brane with angles
in the ten-dimensional supergravity models
 and discussed their applications to 
cosmology and dynamics of branes. 
Our solutions, which have been constructed using the 
T-duality map between the type IIA and type IIB supergravity theories,
  are different from 
the known dynamical intersecting brane solutions 
in supergravity theories. 
These solutions are obtained 
by replacing a constant $c$ in the warp factor $h=c+h_1(z)$ of a 
supersymmetric static solution 
with a linear function of the coordinates 
$x^{\mu}$. This feature is shared by a wide class of supersymmetric 
solutions beyond the examples considered in the present paper, 
In the case of intersecting branes, the field equations normally 
indicate that time dependent solutions in supergravity can be found if only
one harmonic function in the metric depends on time. However,
the solutions of the intersecting brane with angles can contain 
two functions depending on both time as well as overall 
transverse space coordinates. 

We then  construct cosmological models from those solutions 
by smearing some dimensions and compactifying the internal space.
We find the Friedmann-Lema\^itre-Robertson-Walker (FLRW)
 cosmological solutions with power-law expansion.
Unfortunately, the power of the
scale factor is so small that the solutions of field equations cannot 
give a realistic expansion law. 
The fastest expansion power is $\lambda=5/21\approx 0.238$,
which is found in the case of the 
 D3-D5 brane for the U1 type 
and in the D0-D2 brane system for the U2 type. 
This means that we have to include 
additional matter on the brane 
in order to obtain a realistic expanding universe. 
The properties we have discovered would also 
give a clue to investigate 
cosmological models in more complicated setup, such as D-brane 
with angles in the ten-dimensional string theory 
\cite{Breckenridge:1997ar, Ohta:1997ku, Ohta:1997fr, Kitao:1998vn}.

We then discuss the dynamics of branes. 
we have found that 
when the spacetime is contracting in ten dimensions, 
each brane approaches the others as the time evolves.
All domain between branes connected initially ($t<0$), 
but it shrinks as the time increases. 
However, for the D$(p-2)$-D$p$-brane system 
($p\le 7$) without smearing branes,  
a singularity appears before branes collide.
and eventually the topology of the spacetime changes such that 
parts of the branes are separated by a singular region surrounding 
each brane. Thus, the solution
 cannot describe the collision of two branes. 
 In contrast, the D6-D8-brane system or the smeared D$(p-2)$D$p$ brane 
system with one uncompactified extra dimension
can provide an example of colliding branes 
(and  collision of the universes), if they have the same charges.

We also our results in ten-dimensional spacetime and
those in the effective five-dimensional theory.
Although the present models allow the Kaluza-Klein compactification,
i.e., the dynamics is still correct in the effective theory,
the behavior of collision looks different.
The collision in ten dimensions is more violent 
than that in the effective five dimensions. 
It is just because the definitions of the distances are different.
Hence we have to careful to analyze our results obtained in 
the effective theory. 

Although the present examples illustrated in the this paper 
provide neither realistic cosmological model nor
collision of branes (or of the universes), 
the features of the solutions or 
the method to obtain them  could 
open new directions to study how to 
construct a realistic dynamics of branes
as well as an appropriate higher-dimensional 
cosmological model.

\section*{Acknowledgments}
We would like to thank N. Ohta for valuable comments. 
K.U. is grateful to
P. Di Vecchia for discussions. 
The work of K.M. was partially supported by the Grant-in-Aid for Scientific
Research Fund of the JSPS (Grant No.22540291) 
and by the Waseda University Grants for
Special Research Projects. 
K.U. is supported by Grant-in-Aid for Young Scientists (B) of JSPS Research,
under Contract No. 20740147.

\newpage

\appendix

\section{Dynamical brane in massive supergravity}
\label{sec:massive}
In this appendix, we will derive another time dependent solution 
for D6-D8 brane system.

The action for the massive IIA supergravity 
in the Einstein frame can be written as \cite{Singh:2001gt, Singh:2002eu}
\Eqr{
S&=&\frac{1}{2\kappa^2}\int \left(R\ast{\bf 1}
 -\frac{1}{2}d\phi \wedge \ast d\phi
 -\frac{1}{2\cdot 2!}\e^{3\phi/2}F_{(2)}\wedge\ast F_{(2)}
 -\frac{1}{2\cdot 3!}\e^{-\phi}H_{(3)}\wedge\ast H_{(3)}\right.\nn\\
 &&\left.-\frac{1}{2\cdot 4!}\e^{\phi/2}\bar{F}_{(4)}\wedge\ast\bar{F}_{(4)}
 -\frac{1}{2}\e^{5\phi/2}m_0^2\ast {\bf 1}
 -\frac{1}{2}B_{(2)}\wedge \bar{F}_{(4)}\wedge \bar{F}_{(4)}\right)\,,
\label{MIIA:action:Eq}
}
where $\kappa^2$ is the ten-dimensional gravitational constant, 
$m_0$ is constant, $\ast$ is the Hodge dual operator 
in the ten-dimensional spacetime, and $F_{(2)}$, $H_{(3)}$, 
$\bar{F}_{(4)}$ are 2-form, 3-form, 4-form field strength, respectively.
The expectation values of fermionic fields are assumed to vanish.
The field strengths in the action (\ref{MIIA:action:Eq}) are given by
\Eqrsubl{MIIA:strength:Eq}{
H_{(3)}&=&dB_{(2)}\,,\\
F_{(2)}&=&dC_{(1)}+m_0B_{(2)}\,,\\
F_{(4)}&=&\bar{F}_{(4)}+C_{(1)}\wedge H_{(3)}\,,\\
\bar{F}_{(4)}&=&dC_{(3)}+m_0B_{(2)}\wedge B_{(2)}\,,
}
where $C_{(1)}$, $B_{(2)}$, $C_{(3)}$ are 1-form, 2-form, 3-form, 
respectively.


After variations with respect to the metric, the scalar field, 
and the gauge field, the field equations of the D6-D8 brane system 
can be written by
\Eqrsubl{D6D8:equations:Eq}{
&&R_{MN}=\frac{1}{2}\pd_M\phi \pd_N \phi+\frac{1}{16}\e^{5\phi/2}m_0^2
+\frac{1}{2\cdot 2!}\e^{3\phi/2} 
\left(2F_{MA} {F_N}^{A}-\frac{1}{8}g_{MN} F_{(2)}^2\right)\nn\\
&&\hspace{1.5cm}+\frac{1}{2\cdot 3!}\e^{-\phi} 
\left(3H_{MAB} {H_N}^{AB}
-\frac{1}{4}g_{MN} H_{(3)}^2\right)\,,
   \label{D6D8:Einstein:Eq}\\
&&d\ast d\phi=\frac{3}{4\cdot 2!}
\e^{3\phi/2}F_{(2)}\wedge \ast F_{(2)}
-\frac{1}{2\cdot 3!}\e^{-\phi}H_{(3)}\wedge \ast H_{(3)}
+\frac{1}{2}\e^{5\phi/2}m_0^2\ast{\bf 1}\,,
   \label{D6D8:scalar:Eq}\\
&&d\left(\e^{3\phi/2}\ast F_{(2)}\right)=0\,,
   \label{D6D8:F:Eq}\\
&&d\left(\e^{-\phi}\ast H_{(3)}\right)+m_0\,\e^{3\phi/2}\ast F_{(2)}=0\,.
   \label{D6D8:H:Eq}
}
To solve the field equations, we assume the ten-dimensional metric
in the form
\Eqr{
ds^2&=& h_\theta^{1/4}(x, z)h^{7/8}(x, z)\left[
      h^{-1}(x, z)q_{\mu\nu}(\Xsp)dx^{\mu}dx^{\nu}
     +h_\theta^{-1}(x, z)u_{ij}(\Ysp)dy^idy^j
+dz^2\right]\,,~~~~
   \label{D6D8:metric:Eq}
}
where $q_{\mu\nu}$ is a seven-dimensional metric which
depends only on the seven-dimensional coordinates $x^{\mu}$, 
and $u_{ij}$ is the two-dimensional metric which
depends only on the two-dimensional coordinates $y^i$, and the 
function $h_\theta$ is given by
\Eq{
h_\theta=1+\cos^2\theta\left(h-1\right)\,.
   \label{D6D8:h5:Eq}
 }
The metric form (\ref{D6D8:metric:Eq}) is
a straightforward generalization of the case of a static D6-D8 brane
system with a dilaton coupling \cite{Singh:2002eu}.
Furthermore, we assume that the scalar field $\phi$, the parameter 
$m_0$ and the gauge field strengths are given by
\Eqrsubl{D6D8:fields:Eq}{
\e^{\phi}&=&h_\theta^{-1/2}h^{-3/4},\\
B_{(2)}&=&\tan\theta\left(h_\theta^{-1}-1\right)\Omega(\Ysp)\,,\\
dC_{(1)}&=&m\sin\theta\,\Omega(\Ysp)\,,\\
m&=&m_0\left(\cos\theta\right)^{-1}\,,
      }
where $\Omega(\Ysp)$ denotes the volume 2-form,
\Eqr{
\Omega(\Ysp)&=&\sqrt{u}\,dy^1\wedge dy^2\,.
}
The assumptions on the ten-dimensional metric and fields 
correspond to the following brane configuration:
{\scriptsize
\begin{center}
\begin{tabular}{|c|c|c|c|c|c|c|c|c|c|c|}
\hline
&0&1&2&3&4&5&6&7&8&9\\
\hline
D6 & $\circ$ & $\circ$ & $\circ$ & $\circ$ & $\circ$ & $\circ$ & $\circ$ 
&&& \\
\hline
D8 & $\circ$ & $\circ$ & $\circ$ & $\circ$ & $\circ$ & $\circ$ 
& $\circ$ & $\circ$ & $\circ$ & \\ 
\hline
$X^A$ & $~t~$ & $x^1$ & $x^2$ & $x^3$ & $x^4$ & $x^5$ & $x^6$
& $y^1$ & $y^2$ & $~z~$ \\
\hline
\end{tabular}
\end{center}
}
\noindent
Let us first consider the Einstein equations 
(\ref{D6D8:Einstein:Eq}). 
Using the assumptions (\ref{D6D8:metric:Eq}) and 
(\ref{D6D8:fields:Eq}), the Einstein equations are given by
\Eqrsubl{D6D8:cEinstein:Eq}{
&&\hspace{-0.9cm}R_{\mu\nu}(\Xsp)-h^{-1}D_{\mu}D_{\nu}h
+\frac{1}{16}q_{\mu\nu}h^{-1}\left(\Box_{\Xsp}h
+h^{-1}\pd_z^2h\right)-\frac{1}{8}q_{\mu\nu}h_\theta^{-1}
\left(\Box_{\Xsp}h_\theta+h^{-1}\pd_z^2h_\theta\right)\nn\\
&&
+\frac{1}{16}h^{-3}h_\theta^{-1}\left(h_\theta-2\sin^2\theta\right)
\left[\left(\pd_z h\right)^2-m^2\right]=0\,,
 \label{D6D8:cEinstein-mu:Eq}\\
&&\hspace{-0.9cm}\pd_{\mu}\pd_z h=0\,,
 \label{D6D8:cEinstein-mz:Eq}\\ 
&&\hspace{-0.9cm}R_{ij}(\Ysp)
-\frac{7}{16}u_{ij}h_\theta^{-1}\left(\Box_{\Xsp}h
+h^{-1}\pd_z^2h\right)+\frac{3}{8}u_{ij}hh_\theta^{-1}
\left(\Box_{\Xsp}h_\theta+h^{-1}\pd_z^2h_\theta\right)\nn\\
&&
+\frac{1}{16}h^{-2}h_\theta^{-3}\left(h_\theta^3+6\sin^2\theta\right)
\left[\left(\pd_z h\right)^2-m^2\right]=0\,,
 \label{D6D8:cEinstein-ij:Eq}\\
&&\hspace{-0.9cm}-\frac{7}{16}
\left(\Box_{\Xsp}h+h^{-1}\pd_z^2h\right)
-\frac{1}{8}hh_\theta^{-1}
\left(\Box_{\Xsp}h_\theta+h^{-1}\pd_z^2h_\theta\right)
+\frac{1}{16}h^{-2}\left[\left(\pd_z h\right)^2-m^2\right]=0\,,
 \label{D6D8:cEinstein-zz:Eq} 
}
where $D_{\mu}$ is the covariant derivative with respective to 
the metric $q_{\mu\nu}$, $\Box_{\Xsp}$ is
the Laplace operator on the space of 
$\Xsp$, and $R_{\mu\nu}(\Xsp)$, $R_{ij}(\Ysp)$ 
are the Ricci tensors of the metrics $q_{\mu\nu}(\Xsp)$, 
$u_{ij}(\Ysp)$, respectively.

From \Eqref{D6D8:cEinstein-mz:Eq}, 
the warp factor $h$ must be in the form
\Eq{
h(x, z)= K_0(x)+K_1(z)\,.
  \label{D6D8:warp:Eq}
}
With this form of $h$, the other components of
the Einstein equations (\ref{D6D8:cEinstein:Eq}) 
are rewritten as
\Eqrsubl{D6D8:c1Einstein:Eq}{
&&\hspace{-1.2cm}R_{\mu\nu}(\Xsp)-h^{-1}D_{\mu}D_{\nu}K_0
+\frac{1}{16}q_{\mu\nu}h^{-1}\left(\Box_{\Xsp}K_0
+h^{-1}\pd_z^2K_1\right)-\frac{1}{8}q_{\mu\nu}h_\theta^{-1}
\left(\Box_{\Xsp}K_0+h^{-1}\pd_z^2K_1\right)\nn\\
&&
+\frac{1}{16}h^{-3}h_\theta^{-1}\left(h_\theta-2\sin^2\theta\right)
\left[\left(\pd_z K_1\right)^2-m^2\right]=0\,,
 \label{D6D8:c1Einstein-mu:Eq}\\
&&\hspace{-1.2cm}R_{ij}(\Ysp)
-\frac{7}{16}u_{ij}h_\theta^{-1}\left(\Box_{\Xsp}K_0
+h^{-1}\pd_z^2K_1\right)+\frac{3}{8}\cos^2\theta u_{ij}hh_\theta^{-1}
\left(\Box_{\Xsp}K_0+h^{-1}\pd_z^2K_1\right)\nn\\
&&
+\frac{1}{16}h^{-2}h_\theta^{-3}\left(h_\theta^3+6\sin^2\theta\right)
\left[\left(\pd_z K_1\right)^2-m^2\right]=0\,,
 \label{D6D8:c1Einstein-ij:Eq}\\
&&\hspace{-1.2cm}-\frac{7}{16}\left(\Box_{\Xsp}K_0+h^{-1}\pd_z^2K_1\right)
-\frac{1}{8}hh_\theta^{-1}
\left(\Box_{\Xsp}K_0+h^{-1}\pd_z^2K_1\right)
+\frac{1}{16}h^{-2}\left[\left(\pd_z K_1\right)^2-m^2\right]=0\,.~~~
 \label{D6D8:c1Einstein-zz:Eq}
}
Let us next consider the 2-form field $B_{(2)}$ and 1-form $A_{(1)}$.
Under the assumption (\ref{D6D8:fields:Eq}), 
the equation of motion for the gauge field $B$ becomes 
\Eqr{
\sin2\theta\left[\left(\Box_{\Xsp} K_0
+h^{-1}\pd_z^2 K_1\right)-h^{-2}\left\{\left(\pd_zK_1\right)^2
-m^2\right\}\right]\Omega(\Xsp)\wedge dz=0\,,
   \label{D6D8:H3:Eq}
 }
where we used Eq.~(\ref{D6D8:warp:Eq}).
Then, for $\sin2\theta\ne 0$, Eq.~(\ref{D6D8:H3:Eq}) is reduced to
\Eq{
\Box_{\Xsp}K_0=0\,, ~~~ \pd_z K_1=\pm m\,.
   \label{D6D8:warp2:Eq}
}
\Eqref{D6D8:warp2:Eq} thus gives
\Eq{
K_1(z)=1\pm mz\,.
  \label{D6D8:L:Eq}
}
\Eqref{D6D8:H3:Eq} is automatically satisfied for $\sin2\theta=0$. 
Let us next consider the scalar field equation.
Substituting the forms of the function $h$ (\ref{D6D8:warp:Eq}) 
into the equation of motion for the scalar field  
(\ref{D6D8:scalar:Eq}), we obtain
\Eqr{
&&\left(5h_\theta-2\sin^2\theta\right)\left[h^2\Box_{\Xsp}K_0
+h_\theta\left\{\left(\pd_z K_1\right)^2-m^2\right\}\right]=0\,.
   \label{D6D8:scalar3:Eq}
}
Thus, the warp factor $h$ should
satisfy  Eq.~(\ref{D6D8:warp2:Eq}). 
Then, the Einstein equations reduce to  
\Eqrsubl{D6D8:Einstein equations:Eq}{
&&R_{\mu\nu}(\Xsp)=0\,,
   \label{D6D8:Ricci tensor mu:Eq}\\
&&R_{ij}(\Ysp)=0\,,
   \label{D6D8:Ricci tensor ij:Eq}\\ 
&&h(x, z)=K_0(x)+K_1(z)\,;~~
D_{\mu}D_{\nu}K_0=0\,,~~K_1(z)=1\pm mz\,.
   \label{D6D8:warp factor h6:Eq}
 }      
As a special example, we consider the case
\Eq{ 
q_{\mu\nu}=\eta_{\mu\nu}\,,
\quad u_{ij}=\delta_{ij}\,,
 \label{D6D8:s-metric:Eq}
 }
where $\eta_{\mu\nu}$ is the seven-dimensional 
Minkowski metric and $\delta_{ij}$ is 
the two-dimensional Euclidean metric. 
In this case, the solution for $h$ can be obtained
explicitly as
\Eq{ 
h(x, z)=c_{\mu}x^{\mu}\pm m(z-z_0)\,,
 \label{D6D8:h6:Eq}
}
where $c_{\mu}$ and $z_0$ are constant parameters. 


Here we shall discuss the case of $p=8$.
It provides us a colliding-brane model in a massive supergravity.
It may capture the essence of brane collision.
The dynamical D8-brane solution is written as
\Eqr{
ds^2&=&\cos^{1\over 2}\theta\left[c_0t+\sum_{l}M_l|z-z_l|\right]^{7/8}
\left[c_0t+\sum_{l}M_l
|z-z_l|+\tan^2\theta\right]^{1/4}\nn\\
&&\times 
\left[{1\over c_0t+\sum_{l}M_l|z-z_l|}\,
\eta_{\mu\nu}dx^{\mu}dx^{\nu}\right.\nn\\
&&\left.~~~
+{1\over\cos^2\theta \left[
c_0t+\sum_{l}M_l|z-z_l|+\tan^2\theta\right]}\,
\delta_{ij}dy^idy^j+\delta_{ab}dz^adz^b\right]\,,
   \label{D8:metric:Eq}
}
where the constant $\,z_l$ denotes the position of the D8-brane with
charge $M_l$.

Let us consider the two D8-branes with the brane charge $M_1$ at $z=0$
and the other $M_2$ at $z=L$.
The proper distance between two D8-branes is given by
\Eqr{
d(t)=\cos^{1\over 4}\theta\int^L_0 dz \left(c_0t+
 M_1|z|+M_2|z-L|\right)^{\frac{7}{16}}
 \left(c_0t+
 M_1|z|+M_2|z-L|+\tan^2\theta\right)^{\frac{1}{8}}
\,.~~~~
  \label{D8:length:Eq}
}

\begin{figure}[h]
 \begin{center}
  \includegraphics[keepaspectratio, scale=0.5]{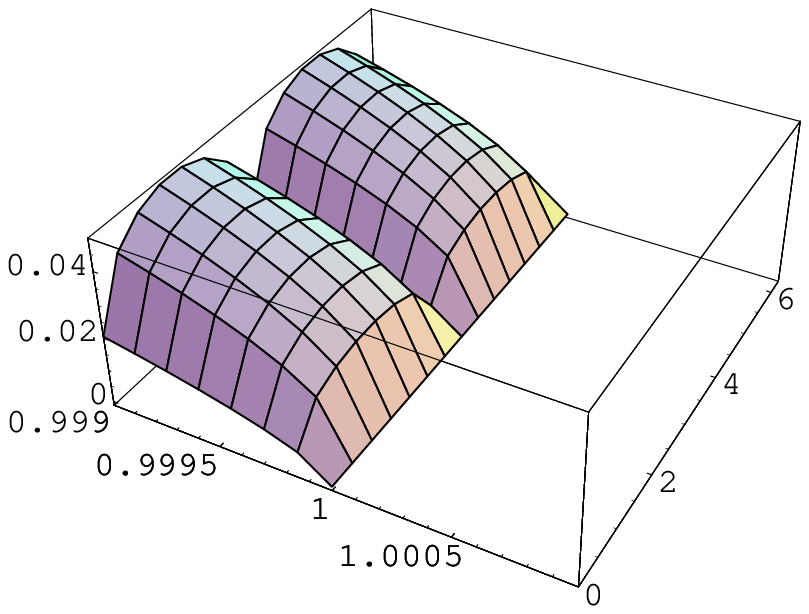}
\put(-120, 70){$d(t)$}
\put(-70, 0){$t$}
\put(-5,20){$\theta$}
  \hskip 3cm
      \includegraphics[keepaspectratio, scale=0.5]{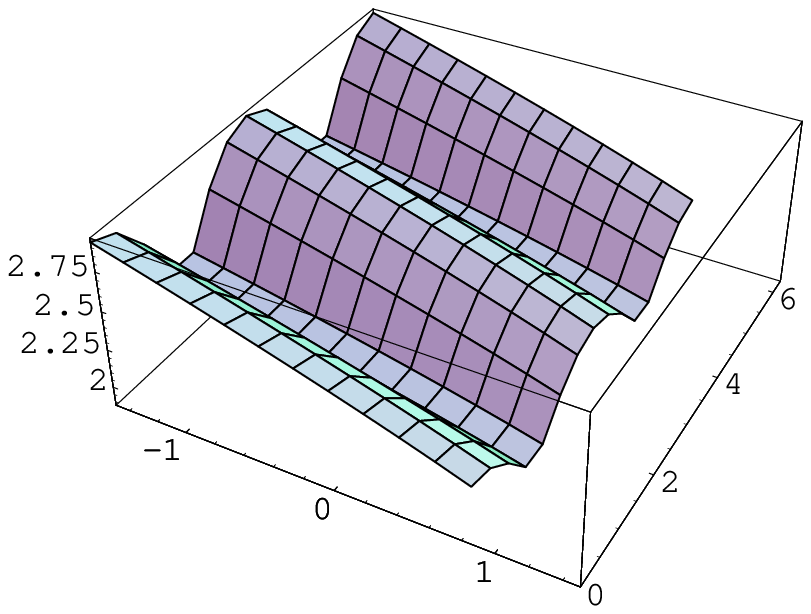}
\put(-120, 70){$d(t)$}
\put(-70, 0){$t$}
\put(-5, 20){$\theta$}
\\
(a) $M_1=M_2=1$\hskip 3.5cm
(b) $M_1=10$, $M_2=1$\\
  \caption{\baselineskip 14pt
The proper distance given in 
  \eqref{D8:length:Eq} is depicted. We fix $c_0=-1$ and $L=1$. 
  The proper distance decreases as $t$ increases. 
  If two D8-brane satisfy $M_1=M_2$, 
  it causes the complete collision at $t=1$ simultaneously. 
  For $M_1\ne M_2$, a singularity appears at $t=t_s>0$ when the distance 
  is still finite. Then, the solution does not describe 
the collision of two D8-branes.}
  \label{fig:ma}
 \end{center}
\end{figure}
For $c_0<0$,  the proper distance decreases as
$t$ increases, and it eventually vanishes at $t=-ML/c_0$ if two brane charges 
are equal such that $M_1=M_2=M$.
Hence, one D8-brane approaches the other as time progresses,
causing the complete collision at $t=-ML/c_0$. 
We note that the collision occurs simultaneously. 
This behavior, however, changes if  $M_1\neq M_2$. 
A singularity forms at $t=t_s\equiv-[M_1|z|+M_2|z-L|]/c_0>0$, 
when the distance is still finite. 
We show $d(t)$ in Fig. \ref{fig:ma}.

For $t<0$, each brane gradually
separates from the other as the time goes in the past.

\section{D0-D2-D2-D4 brane system}
\label{sec:D0D2D4}

In this appendix, we discuss the solution for involving 
more than two types of D-branes. This is given by the procedure 
of delocalization, rotation and T duality with respect to 
more than one of the transverse coordinates of the original D-brane 
solutions. Let us consider the D0-D2-brane with D4-brane system. 
The ten-dimensional metric is given by \cite{Breckenridge:1996tt}
\Eqr{
ds^2&=& h^{1/8}(t, z)h_\theta^{1/4}(t, z)h_\psi^{1/4}(t, z)\left[-h^{-1}(t, z)
dt^2
+h_\theta^{-1}(t, z)\gamma_{ij}(\Ysp_1)dy^idy^j\right.\nn\\
     & &\left.+h_\psi^{-1}(t, z)v_{mn}(\Ysp_2)d\eta^{m}d\eta^{n}
     +u_{ab}(\Zsp)dz^adz^b\right]\,,
   \label{D024:metric:Eq}
     }
where $\gamma_{ij}$ is the two-dimensional metric which
depends only on the two-dimensional coordinates $y^i$, 
$v_{mn}$ is the two-dimensional metric which 
depends only on the two-dimensional coordinates $\eta^m$, 
and $u_{ab}$ is the five-dimensional metric which
depends only on the five-dimensional coordinates $z^a$, and the 
functions $h_\theta$, $h_\psi$ are given by
\Eq{
h_\theta(t, z)=1+ \cos^2\theta\left(h-1\right)\,,~~~~
h_\psi(t, z)=1+ \cos^2\psi\left(h-1\right)\,.
   \label{D024:f:Eq}
 }
The assumptions on the ten-dimensional metric and fields 
correspond to the following brane configuration:
{\scriptsize
\begin{center}
\begin{tabular}{|c|c|c|c|c|c|c|c|c|c|c|}
\hline
&0&1&2&3&4&5&6&7&8&9\\
\hline
D0 & $\circ$ & & &&&&&&& \\
\hline
D2 & $\circ$ & $\circ$ & $\circ$ &&&&&&& \\
\hline
D2 & $\circ$ & & & $\circ$ & $\circ$ &
&& & & \\ 
\hline
D4 & $\circ$ & $\circ$ & $\circ$ & $\circ$ & $\circ$ &
&& & & \\ 
\hline
$X^A$ & $~t~$ & $y^1$ & $y^2$ & $\eta^1$ & $\eta^2$ & $z^1$ & $z^2$
& $z^3$ & $z^4$ & $z^5$ \\
\hline
\end{tabular}
\end{center}
}
\noindent
The scalar field $\phi$ and 
the gauge field strengths are given by
\Eqrsubl{D024:fields:Eq}{
\e^{\phi}&=&h^{3/4}\left(h_\theta h_\psi\right)^{-1/2}\,,\\
C_{(1)}&=&\pm\sin\theta\sin\psi\left(h^{-1}-1\right)dt\,,\\
C_{(3)}&=&\pm \cos^{-1}\theta \sin\psi
\left(h_\theta^{-1}-1\right)dt\wedge\Omega(\Ysp_1)\nn\\
&&\pm \cos^{-1}\psi \sin\theta
\left(h_\psi^{-1}-1\right)dt\wedge\Omega(\Ysp_2)
+\omega_{(3)}\,,\\
B_{(2)}&=&\tan\theta\,\left(h_\theta^{-1}-1\right)\Omega(\Ysp_1)
+\tan\psi\,\left(h_\psi^{-1}-1\right)\Omega(\Ysp_2)\,,
}
where $\Omega(\Ysp_1)$ and $\Omega(\Ysp_2)$ denote the volume form,
\Eq{
\Omega(\Ysp_1)=\sqrt{\gamma}\,dy^1\wedge dy^2\,,~~~~
\Omega(\Ysp_2)=\sqrt{v}\,d\eta^1\wedge d\eta^2\,,
}
and the three form $\omega_{(3)}$ satisfies
\Eq{
d\omega_{(3)}=\pm \cos\theta\,
\cos\psi\,\pd_ah\ast_{\Zsp}\left(dz^a\right)\,.
}
Here $\ast_{\Zsp}$ denotes the Hodge operator on $\Zsp$.

Performing the same procedure as in the previous section, 
we find that the field equations are reduced to 
\Eqrsubl{D024:Einstein equations:Eq}{
&&R_{ij}(\Ysp_1)=0\,,
   \label{D024:Ricci tensor ij:Eq}\\
&&R_{mn}(\Ysp_2)=0\,,
   \label{D024:Ricci tensor mn:Eq}\\
&&R_{ab}(\Zsp)=0\,,
   \label{D024:Ricci tensor pq:Eq}\\ 
&&h(t, z)=K_0(t)+K_1(z);~~~\pd_t^2K_0=0\,,~~~\lap_{\Zsp}K_1=0\,,
   \label{D024:warp factor h:Eq}\\
&&h_\theta(t, z)=1+\cos^2\theta(h-1)\,,~~~~h_\psi(t, z)=1+\cos^2\psi(h-1)\,,
   \label{D024:warp factor h5:Eq} 
 }      
where $\triangle_{\Zsp}$ is 
the Laplace operators on the space of $\Zsp$, and 
$R_{ij}(\Ysp_1)$, $R_{mn}(\Ysp_2)$, 
and $R_{ab}(\Zsp)$ are the Ricci tensors
of the metrics $\gamma_{ij}(\Ysp_1)$, $v_{mn}(\Ysp_2)$, 
and $u_{ab}(\Zsp)$, respectively.

Let us consider the case $u_{ab}=\delta_{ab}$ in more detail, 
where $\delta_{ab}$ are the five-dimensional Euclidean metric.
In this case, a solution for the warp factor $h$ 
can be obtained explicitly as
\Eqr{ 
h(t, z)=ct+\tilde{c}+\sum_l\frac{M_l}{|z^a-z^a_l|^3}\,,
 \label{D024:warp:Eq}
}
where $c$, $\tilde{c}$, $M_l$ and $z_l$ are constant parameters.

If the branes exist at the origin of $\Zsp$ space, 
introducing a radial coordinate $r$ by
\Eq{
\delta_{ab}dz^adz^b=dr^2+r^2d\Omega^2_4\,,
}
we find that the function $h$ is expressed as  
\Eq{
h=c_0t+\tilde c+\left(\frac{m}{r}\right)^3\,,
}
where $d\Omega^2_4$ is the line element of four-dimensional 
sphere, and $c_0$, $\tilde c$, and $m$ are constants. 
In the limit $r\rightarrow 0$, 
the metric \eqref{D024:metric:Eq} gives
\Eqr{
\hspace{-0.1cm}
ds^2&=&\left(\cos\theta\cos\psi\right)^{1/2}
\left(\frac{m}{r}\right)^{-1/8}
\left[\left(\frac{m}{r}\right)^{-1}
\left(-dt^2+\cos^{-2}\theta\gamma_{ij}dy^idy^j
+\cos^{-2}\psi v_{mn}d\eta^md\eta^n\right)\right.\nn\\
&&\left. +m^2\frac{dr^2}{r^2}+m^2d\Omega_4^2\right]\,,
   \label{D024:metric2:Eq}
}
while the dilaton is given by
\Eq{
\lim_{r\rightarrow 0}\e^{\phi}=\left(\cos\theta\cos\psi\right)^{-1}
\left(\frac{m}{r}\right)^{-3/4}\,.
    \label{D024:dilaton2:Eq}
}
Hence the ten-dimensional 
metric with $\theta=0$, $\psi=0$, $\gamma_{ij}=\delta_{ij}$, and 
$v_{mn}=\delta_{mn}$ 
becomes a warped ${\rm AdS}_6\times {\rm S}^4$ spacetime.

The dynamical solution can be obtained by the same procedure of 
the delocalization and rotation on a D2-brane. 
Let us single out two orthogonal planes $(y^1, y^2)$ and $(\eta^1, \eta^2)$.
If we apply the procedure of the delocalization and rotation 
on a D2-brane with respect to the $(y^1, y^2)$ plane, 
followed by T-duality map (\ref{D4:duality:Eq}), 
we can obtain the solution for a D3-D1 brane, 
where the rotation angle is given by $\theta$. 
After repeating the same procedure of the delocalization and rotation 
on a D3-brane with respect to the $(\eta^1, \eta^2)$ plane 
- rotating by $\psi$ 
to $(\eta^1, \eta^2)$ - , followed by T-duality map 
\cite{Bergshoeff:1995as, Breckenridge:1997ar}
\Eqr{
&&g^{(\rm A)}_{yy}=\frac{1}{g^{(\rm B)}_{yy}}\,,~~~~
g^{(\rm A)}_{\mu\nu}=g^{(\rm B)}_{\mu\nu}
-\frac{g^{(\rm B)}_{y\mu}g^{(\rm B)}_{y\nu}
-B^{(\rm B)}_{y\mu}B^{(\rm B)}_{y\nu}}
{g^{(\rm B)}_{yy}}\,,~~~~
g^{(\rm A)}_{y\mu}=-\frac{B^{(\rm B)}_{y\mu}}{g^{(\rm B)}_{yy}}\,,\nn\\
&&\e^{2\phi_{(\rm A)}}=\frac{\e^{2\phi_{(\rm B)}}}{g^{(\rm B)}_{yy}}\,,~~~~
C_{\mu}=C_{y\mu}+C_{(0)}B_{y\mu}^{(\rm B)}\,,~~~~C_y=-C_{(0)}\,,\nn\\
&&B^{(\rm A)}_{\mu\nu}=B^{(\rm B)}_{\mu\nu}
+2\frac{B^{(\rm B)}_{y[\mu}\,g^{(\rm B)}_{\nu]y}}{g^{(\rm B)}_{yy}}\,,~~~~
B^{(\rm A)}_{y\mu}=-\frac{g^{(\rm B)}_{y\mu}}{g^{(\rm B)}_{yy}}\,,~~~~
C_{y\mu\nu}=C_{\mu\nu}
+2\frac{C_{y[\mu}\,g^{(\rm B)}_{\nu]y}}{g^{(\rm B)}_{yy}}\,,\nn\\
&&C_{\mu\nu\rho}=C_{\mu\nu\rho y}+\frac{3}{2}\left(
C_{y[\mu}\,B^{(\rm B)}_{\nu\rho]}
-B^{(\rm B)}_{y[\mu}\,C_{\nu\rho]}
-4\frac{B^{(\rm B)}_{y[\mu}\,C_{|y|\nu}g^{(\rm B)}_{\rho]y}}
{g^{(\rm B)}_{yy}}\right)\,,
   \label{D1:duality:Eq}
}
we can construct the solution of the D0-D2-D4-brane system 
(\ref{D024:warp:Eq}) \cite{Breckenridge:1996tt}.
Here $y$ is the coordinate to which the T dualization is applied, and 
$\mu$, $\nu$, and $\rho$ denote the coordinates other than $y$.

By smearing $\Ysp_1$ space, $\Ysp_2$ space, and some of $\Zsp$ space (
$0\leq d_\Zsp\leq 2$ dimensions)and compactifying them, 
we can construct the type U2 isotropic and homogeneous three space
as our universe.
We can also discuss collision of branes (or universes).

\newpage



\end{document}